\documentstyle[epsfig,amsmath,times]{mn}
\def\spose#1{\hbox to 0pt{#1\hss}}

\def\oo{[O\textsc{ii}]\,}
\def\lta{\mathrel{\spose{\lower 3pt\hbox{$\mathchar"218$}}\raise 2.0pt\hbox{$\mathchar"13C$}}}
\def\gta{\mathrel{\spose{\lower 3pt\hbox{$\mathchar"218$}}\raise 2.0pt\hbox{$\mathchar"13E$}}}
\def\arcsec{$^{\prime\prime}$}

\title[HST and UKIRT imaging observations - I]
{HST and UKIRT imaging observations of $\mathbf{z \sim 1}$ 6C radio
galaxies - I. The data}
      
\author[K.\,J.\, Inskip {\it et al.}]
{K.\,J.\, Inskip$^1$\footnotemark, P.\,N.\,Best$^2$,
M.\,S.\,Longair$^1$, S.\,Rawlings$^3$, H.\,J.\,A.\,R\"{o}ttgering$^4$
and \cr S.\,Eales$^5$\\ 
$^1$ Cavendish Laboratory, Madingley Road, Cambridge, CB3 0HE,\\ $^2$
Institute for Astronomy, Royal Observatory Edinburgh, Blackford Hill,
Edinburgh, EH9 3HJ\\ $^3$ Department of Astrophysics, University of
Oxford, Keble Road, Oxford OX1 3RH\\ $^4$ Sterrewacht Leiden, Postbus
9513, 2300 RA Leiden, the Netherlands\\ $^5$ Department of Physics and
Astrophysics, University of Wales Cardiff, PO Box 913, Cardiff CF2 3YB\\}

\date{ }

\pagerange{\pageref{firstpage}--\pageref{lastpage}}

\pubyear{2003}

\begin{document}

\label{firstpage}
\maketitle

\begin{abstract}

The results of Hubble Space Telescope and UKIRT imaging observations
are presented for a sample of 11 6C radio galaxies with redshifts
$0.85 < z < 1.5$.  The observations of the 6C sources reveal a variety
of different features, similar to those observed around the higher
power 3CR sources of similar redshifts. However, the extent and
luminosity of the aligned emission appears less extreme in the case of
the 6C radio galaxies.  For both samples, the aligned emission clearly
cannot be explained by a single emission mechanism; line emission and
related nebular continuum emission, however, often provide a
significant contribution to the aligned emission.
\end{abstract}


\section{Introduction}
\footnotetext{E-mail: kji@mrao.cam.ac.uk} Powerful distant radio
sources (up redshifts of at least unity) are usually hosted by giant elliptical galaxies.  $K$-band
observations of these galaxies show them to have de Vaucouleurs radial
profiles, with the emission dominated by light from the old
stellar populations (e.g. Best, Longair \& R\"{o}ttgering 1998).
However, observations at shorter wavelengths frequently show the host
galaxies to be surrounded by extensive regions of UV/optical continuum and line
emission, particularly at redshifts $z > 0.3$.  At higher redshifts
($z \gta 0.6$) these regions of extended emission are usually observed
to be closely aligned with the radio source axis (e.g. Chambers, Miley
\& van Breugel 1987; McCarthy et al 1987); this is known as the {\it
Alignment Effect}.  

Many different emission mechanisms have been proposed in order to
account for the alignment effect; the most likely options include
extended line emission and nebular continuum radiation (Dickson et al
1995), scattering of the UV continuum from the AGN (e.g. Tadhunter et
al 1992; Cimatti et al 1993) and young stars produced in a radio jet
induced starburst (McCarthy et al 1987).  Generally, no single
mechanism can  account for all the excess emission forming the
alignment effect.  The relative contributions of each of these
processes depend on several factors, including the available gas mass
in the IGM, and the power of the radio source.  Additionally, the
presence of shocks associated with the expanding radio source
(evidence for which has been observed in the emission line spectra of
many radio sources, e.g Clark et al 1998; Best, R\"{o}ttgering \&
Longair 2000; Sol\'{o}zano-I\~{n}arrea, Tadhunter \& Axon 2001; Inskip
et al 2002a) can also influence the continuum alignment effect.  The
passage of radio source shocks  through the cool dense gas clouds can
induce star formation (e.g. Dey et al 1997; Bicknell et al 2000).
Ionizing photons associated with the shocks can boost the luminosity
of certain emission lines, and cause  increased nebular continuum
emission.  The passage of a fast shock can also potentially cause the
break-up of optically thick clouds (Bremer, Fabian \& Crawford 1997;
Mellema, Kurk \& R\"{o}ttgering 2002), increasing the covering factor
for scattering of the UV flux from the AGN.  The more numerous,
smaller clouds will also have a larger cross section for ionization by
the AGN, leading to an increase in the total flux of line emission.
However, although the presence of shocks can enhance the alignment
effect, with the exception of jet-induced star formation shocks are
not necessarily required.

Studies of 3CR radio galaxies (e.g. McCarthy, Spinrad \& van Breugel
1995; Best, Longair \& R\"{o}ttgering 1997) have shown that the
regions of extended optical/UV continuum emission surrounding the
higher redshift sources in the sample are more extensive and better
aligned with the radio source axis than for sources at lower
redshifts.  This may reflect a real evolutionary trend: changes in
radio galaxy environment with cosmic epoch, particularly relating to
the availability of cool dense gas clouds in the local IGM, could
clearly affect the observed alignment effect.  However, the observed
variation in the alignment effect with redshift for 3CR sources may
instead be due to a number of selection effects, including the
rest-frame wavelength of the observed emission, the size of the radio
source and the radio source power.  We now discuss each of these
selection effects in turn.

\begin{enumerate}
\item {\it Rest frame wavelength.}  For observations at a given
wavelength of a sample of galaxies covering a large redshift range,
the rest-frame wavelength at which the galaxies are observed becomes
longer at lower redshifts.  As discussed above, at longer wavelengths
the host galaxy dominates the light, and the alignment effect is
weaker (Best, Longair \& R\"{o}ttgering 1997; Rigler et al
1992).  Additionally, any
wavelength dependence of the alignment effect will express itself as a
change in observed aligned emission properties with redshift. 

\item {\it Radio source size.}  Radio galaxies at high redshift are
observed to have typically smaller projected linear sizes than those
nearby (e.g. Wardle \& Miley 1974). In part, this is argued to be due
to a selection bias within flux limited samples (e.g. Blundell,
Rawlings \& Willott 1999, Neeser et al 1995) caused by a combination
of Malmquist bias and the decrease in radio luminosity with radio size
for individual sources (Kaiser, Dennet-Thorpe \& Alexander 1997).
This change in mean size with redshift is potentially important, as
many aligned emission properties 
show a strong variation with the size of the radio source.  Best et al
(1997) found that the emission surrounding the smaller radio galaxies
in their sample was usually brighter and more extensive than that seen
around the larger radio sources in the sample, and was better aligned
with the radio axis, leading to an enhanced alignment effect.  The 
effects of shocks are particularly pronounced in the case of  smaller
radio sources (i.e. those with a projected physical size of $<
120$kpc) and can lead to increased emission from the extended
structures in these smaller sources.  Therefore, at high redshifts,
the larger number of radio sources with sizes comparable to the
optical host galaxy sizes (or extended emission line region sizes)
will increase the predominance of the alignment effect.
However, this alone cannot account for the apparent redshift evolution of
the alignment effect. 

\item {\it Radio power.}
One final problem with studies of flux limited samples of radio 
galaxies (such as the 3CR sample) is that the
radio power of these sources increases with redshift, leading to a
tight coupling between redshift and radio power.  It is therefore unclear
whether the observed variation in the alignment effect with redshift
reflects a link between radio power and the strength of the observed
alignment effect, or an evolutionary trend with cosmic epoch. 
\end{enumerate}

To improve our understanding of the relative contributions
of different emission mechanisms to the alignment effect,
the influence of radio power and redshift needs to be investigated independently.
The 6C sample of radio galaxies (Eales 1985) provide a population of
radio galaxies ideally suited for this investigation, as these
galaxies are roughly a factor of 6 lower in radio power than 3CR
sources at the same redshift.  We have carried out a program of
multiwavelength imaging and spectroscopic observations of a subsample
of 11 6C radio sources at $z \sim 1$, which are well matched to the
3CR subsample previously studied by Best et al (1997, 1998, 2000).
The results of these observations can then be contrasted with those of
similar observations of 3CR sources matched in either redshift or
radio power.  By comparing Hubble Space Telescope (HST) and UKIRT
observations of 6C and  3CR sources at $z \sim 1$, we can investigate
the effect of radio power on the galaxy colours, and therefore the
excess UV emission producing the alignment effect.  These observations
also allow us to study the morphologies of the aligned structures
surrounding distant radio galaxies at lower radio powers, and to
determine whether radio power influences the extent and luminosity of
the alignment effect, or the type of features observed.

The structure of the paper is as follows.  In section 2, the sample
selection, HST and UKIRT observations and the astrometry are
outlined. In section 3 we present an analysis of the contribution of
line and nebular continuum emission to the observed magnitudes. We
present the HST and UKIRT images in section 4, and conclude with a
discussion of the data in section 5.  The full analyses of the
properties of the host galaxy, the alignment effect and the galaxy
colours, and the comparison of these results with a matched sample of
3CR galaxies at $z \sim 1$ are deferred to two subsequent papers in
this series.  Values for the cosmological parameters $\Omega_0=0.3$,
$\Omega_\Lambda=0.7$ and $H_{0}=65\,\rm{km\,s^{-1}\,Mpc^{-1}}$ are
assumed throughout this paper.

\section{Sample Selection and Observations}
This paper focuses on observations of galaxies selected from the
6CER sample\footnote{Up--to--date information for the revised
6CE sample can be found at: http://www-astro.physics.ox.ac.uk/$\sim$sr/6ce.html.}   
(Rawlings, Eales \& Lacy 2001), a revised version of the sample
originally defined by Eales (1985).  This revised sample is complete,
and consists of 59 radio sources with flux densities at 151 MHz which fall
in the range $2.0 \rm{Jy} < S_{151} < 3.93 \rm{Jy}$, and lie in the
region of the sky $08^h20^m < {\rm RA} < 13^h01^m$, $34^{\circ} <
\rm{Dec} < 40^{\circ}$.   Eleven high redshift 6C sources within the
redshift range $0.85 < z < 1.5$ have been selected from this sample,
excluding those sources identified as quasars (Best et al
1999).  Three further sources in this redshift range are not included
in the sample. 6C1123+34 was excluded on the
basis that it would be barely resolvable in the 5GHz radio
observations of the sample.  Additionally, the first estimate of the redshifts of
6C1212+38 and 6C1045+25 wrongly placed them outside the selection
criteria for this sample.  These galaxies were selected as a comparison sample matched in
redshift to the 3CR $z \sim 1$ subsample of Best {\it et al}
(1997).  At these redshifts, 6C radio galaxies are
approximately 6 times less powerful radio sources than 3CR galaxies.  

 
\subsection{HST observations}
The Hubble Space Telescope (HST) observations were made in 1996/8 (6C0943$+$39, 6C1129$+$37,
6C1011$+$36,   6C1017$+$37, 6C1256$+$36) and 2000/1  (6C0825$+$34,
6C1019$+$39, 6C1100$+$35, 6C1204$+$35,  6C1257$+$36, 6C1217$+$36)
under proposals \#6684 and \#8173 respectively.  The instrument used
for these observations was the Wide Field/Planetary Camera 2
(WFPC2). The dimensions of each of the WFPC2 chips are $800 \times
800$ pixels, and the scale of the three WF chips is $\sim
0.09953$\arcsec\ per pixel, giving a total field of view of about $160
\times 160$ arcsec$^2$.   For the five sources observed in 1996/1998,
the F702W filter was used.  For the three high redshift sources ($z >
1.3$) observed  in 2000/2001, observations were made using the F814W
filter.  These filters were chosen to be at the same rest-frame
wavelength as the HST observations of the 3CR $z \sim 1$ sample of
Best, Longair \& R\"{o}ttgering (1997).  For the remaining three
sources observed  in 2000/2001, two filters were used (F606W and
F814W).  As well as covering the same rest-frame wavelength, the use
of two filters  provides colour information, which is particularly
useful when studying the aligned emission.  For the purposes of later
clustering analysis, the target object was positioned towards the
corner of the WF3 chip, at the approximate centre of the WF/PC field.
 
Observations in each filter consisted of 2-6 individual exposures of
1200s duration.  All exposures were individually calibrated using the
standard Space Telescope Science Institute (STScI) pipeline
processing.  After this stage, images made in the same filter for a
given object were combined using the STSDAS IRAF package
\textsc{gcombine}.  This package combines the data on each WFPC2 chip
by weighted averaging, rejecting most ($\sim 95\%$) of the cosmic rays
in the process.  Any remaining cosmic rays were removed using the IRAF
package  \textsc{cosmicrays}, which rejects pixels above a threshold
level on the basis of their flux  ratio relative to the mean
neighbouring pixel flux.   

The WFPC2 CCDs have a small parallel charge transfer efficiency (CTE)
problem. This correction for this was approximated by Holtzman et al
(1995) as a linear function, whereby pixels values in row 1 remain
unchanged, whilst the pixel values in row 800 need to be increased by a
factor of 4\%. Although a more realistic CTE estimate requires that the
observing date, target pixel location, target counts and background
level are also taken into account (Biretta, Lubin et al 2002, Dolphin
et al 2000), for the setup of the current observations (high
background levels, large aperture photometry) the Holtzman et al
approximation (accurate to within 0.01 magnitudes) is fully consistent
with the Dolphin et al corrections.  Given that the error on this
approximation is substantially smaller than the statistical
photometric errors, the Holtzman et al correction was adopted for the
sake of simplicity. 

Photometry of the radio galaxies was carried out using the IRAF
\textsc{apphot} packages. 9\arcsec\ diameter apertures were used in
order to include most of the light from the host galaxy and  aligned
structures, and the background level was determined from an annulus
surrounding the object aperture.    The background level statistics
were also measured for 30 apertures in empty regions of the sky, in
order to estimate the systematic uncertainty of the background
subtraction due to variations in sky intensity across the field of
view.   For some sources, flux from neighbouring objects lay within
the 9\arcsec\ object apertures.  Where the objects  were well
separated from the radio galaxy, the additional flux could be
estimated and was removed from the aperture counts. For later analysis
of the colours of the target objects, flux  was also measured in a
4\arcsec\ diameter aperture. In this case, any flux contamination by
other objects was generally negligible, and no corrections were
necessary.
 
WFPC2 counts were converted to photometric magnitudes using the method
described by Holtzman {\it et al} (1995): \\
$m = -2.5 {\rm log}_{10} \left(\frac{F}{T \times 1.1 \times 2.006}\right) + M_{ZP}$\\
where $F$ is the measured object counts, $T$ is the exposure time and
$M_{ZP}$ is the zero point magnitude for the filter.  The zero point
magnitudes used are the most recent update\footnote{see
http://www.noao.edu/staff/dolphin/wfpc2\_calib/} based on the data presented
in Dolphin (2000); these are 22.101, 21.674 and 20.855 for the F606W,
F702W and F814W filters respectively.  The factor of 2.006 corresponds
to the required gain conversion for the WF3 chip, since the
photometric calibration data for WFPC2 are observed with a gain of
14$e^-$/DN, whereas these observations were made with a gain of
7$e^-$/DN.  The additional factor of 1.1 is required to account for
aperture correction of observations of photometric standard stars
(Holtzman et al 1995).  The resulting magnitudes have been corrected
for galactic extinction using the $E(B-V)$ for the Milky Way from the
NASA Extragalactic Database (NED) and the parametrized galactic
extinction law of Howarth (1983).
 
There are several sources of uncertainty in the measurement of HST fluxes.
Errors in the photometric calibration include 1\% for the CTE effect,
1\% for the gain ratio correction, 1\% for the photometric aperture
corrections and 2\% from the accuracy of the zero point magnitudes.
Added in quadrature to these are the photometry errors determined
using \textsc{apphot}:
the percentage error due to the
Poisson noise from the detected counts ($1/\sqrt{G \times N_f \times
C}$, where $G$ is the Gain, $N_f$ is the number of frames used in each
\textsc{gcombine}d exposure, and $C$ is the total counts per frame in the
aperture), the percentage random error due to the sky counts in the
aperture ($(\sqrt{N_{pix}} \times \sigma)/C$, where $N_{pix}$ is the
number of pixels in the object aperture, and $\sigma$ is the mean
standard deviation of the sky counts, calculated from 30 different sky
apertures), and the percentage error due to the accuracy to which the
mean sky value can be calculated ($(N_{pix}\times E_{Sky})/C$).  For
this last systematic source of error, the sky error $E_{Sky}$ is
calculated as the standard deviation of the 30 mean sky values
calculated in apertures at different locations on the CCD.  
These errors differ slightly from the \textsc{apphot} values
for a single sky aperture, due to the additional consideration
of systematic errors in the sky subtraction.

All magnitudes quoted in this paper for the HST
observations use HST filters.  Conversions to ground based
$UBVRI$ magnitudes can be determined using the tabulated filter data
in Holtzman {\it et al} (1995), although this would add large
uncertainties to the galaxy colours.  For a rough comparison with
ground based magnitudes, $F814W \approx I + 0.07(V-I)$, $F702W \approx R
- 0.3(V-R)$ and $F606W \approx V - 0.25(V-I)$.  
Full details of the HST observations are displayed in Table 1.
 
\begin{table*}
\caption{Full details for the HST observations. The dates of
observations are given in column 3; those made in 1996/8 were under
proposal \#6684, and those in 2000/1 under proposal \#8173. The filter
details are given in columns 4 and 5, and the exposure times of the
observations are listed in column 6.  Columns 7 and 8 give the
photometric magnitudes and errors determined in a 9\arcsec\ diameter
aperture. These 9\arcsec\ magnitudes have been corrected for flux
contamination by adjacent objects. 4  \arcsec\ diameter aperture
magnitudes and their associated errors are given in columns 9 and
10. The 4\arcsec\ magnitudes do not include any correction for flux
contamination by nearby objects.  All magnitudes have been corrected
for galactic extinction.}
\begin{center} 
\begin{tabular} {lccccccccccc}
{[1]}&{[2]}&{[3]}&{[4]}&{[5]}&{[6]} &{[7]}&{[8]}&{[9]}&{[10]}\\
{Source}&{z}&{Observation}&{Filter}&{Central}&{Exposure} &{9\arcsec\ WFPC}&{Error}&{4\arcsec\ WFPC}&{Error}\\
        &   & {Date}  &        &{Wavelength [\AA]}&{Time [s]} &{Magnitude}&   &{Magnitude}&       \\\hline

6C0825+34 & 1.467 & 17/03/00 & F814W & 7924 & 7200 & 21.65 & 0.15 & 22.10 & 0.17 \\
6C0943+39 & 1.035 & 19/05/96 & F702W & 6862 & 4800 & 21.78 & 0.38 & 22.08 & 0.14 \\
6C1011+36 & 1.042 & 19/03/98 & F702W & 6862 & 4800 & 21.59 & 0.19 & 21.73 & 0.07 \\
6C1017+37 & 1.053 & 14/05/98 & F702W & 6862 & 4800 & 21.53 & 0.16 & 21.77 & 0.06 \\
6C1019+39 & 0.922 & 14/05/00 & F606W & 5934 & 2400 & 21.89 & 0.25 & 22.01 & 0.08 \\
          &       &          & F814W & 7924 & 2400 & 19.86 & 0.10 & 20.07 & 0.04 \\
6C1100+35 & 1.440 & 17/05/00 & F814W & 7924 & 7200 & 21.80 & 0.18 & 21.80 & 0.06 \\
6C1129+37 & 1.060 & 08/05/96 & F702W & 6862 & 4800 & 21.52 & 0.14 & 22.02 & 0.07 \\
6C1204+35 & 1.376 & 11/05/00 & F814W & 7924 & 7200 & 20.79 & 0.29 & 21.30 & 0.10 \\
6C1217+36 & 1.088 & 15/04/01 & F606W & 5934 & 2400 & 22.26 & 0.32 & 22.25 & 0.09 \\
          &       &          & F814W & 7924 & 2400 & 20.59 & 0.12 & 20.89 & 0.05 \\
6C1256+36 & 1.128 & 05/07/98 & F702W & 6862 & 4800 & 22.23 & 0.16 & 22.66 & 0.09 \\
6C1257+36 & 1.004 & 01/07/00 & F606W & 5934 & 2400 & 22.04 & 0.26 & 22.48 & 0.09 \\
          &       &          & F814W & 7924 & 2400 & 20.55 & 0.12 & 20.98 & 0.05 
\end{tabular}		           
\end{center}		           
\end{table*}		           

\subsection{UKIRT observations}

The majority of the UKIRT observations were carried out over four
nights in March 2001 using UFTI (the
UKIRT Fast--Track Imager).  This is a 1--2.5$\mu$m camera with a $1024
\times 1024$ HgCdTe array and a plate scale of 0.091\arcsec\, per
pixel, which gives a field of view of 92\arcsec.   
These observations were made in the $J$, $H$ and $K$ infrared
wavebands for each source; full observational details are tabulated in Table
2. The $K$-band observations were generally of 1-2 hours duration in
total.  $J$-band observations were typically 18-45 minutes, and $H$-band
observations typically 9-18 minutes.  

Service time observations were also taken in order 
to make up for time lost due to poor weather.  Of these observations,
the $H$-band service observations of 6C1204+35, 6C1217+36 and
6C1257+36 were carried out using IRCAM3.  IRCAM3 is a cooled
1--5$\mu$m camera with a $256 \times 256$ InSb array and a plate scale
of 0.081\arcsec\, per pixel, which gives a field of view of
20.8\arcsec.  Additional UFTI observations of 6C1100+35, 6C1129+37 and  
6C1204+35 were kindly obtained in April 2002 by Mairi Brookes and
Philip Best. 

\begin{table*}
\caption{Full details of UKIRT observations for the eleven 6C sources.
The dates of observations are given in column 4. An $^{S}$ indicates
that the observations were made in service mode, $^{B}$ indicates data
were taken by Mairi Brookes and Philip Best, and $^{I}$ denotes that
the observations used IRCAM3 rather than UFTI.  Column 5 lists the
seeing conditions of each set of observations, and column 6 gives the
total useful exposure time in seconds, with the total photometric
exposure time in brackets.  9\arcsec\ diameter aperture $J$, $H$ and
$K$ magnitudes are given in column 7 with their associated error
(column 8).  4\arcsec\ diameter aperture $J$, $H$ and $K$ magnitudes
are given with their associated error in columns 9 and 10.
For the 9\arcsec\ magnitudes only, corrections have been made for flux
contamination from adjacent objects.  All magnitudes have been
corrected for galactic extinction. } 
\begin{center} 
\footnotesize{}
\begin{tabular} {lccccccccc}
{[1]}&{[2]}&{[3]}&{[4]}&{[5]}&{[6]} &{[7]}&{[8]}&{[9]}&{[10]}\\
{Source}&{z}&{Filter}& {Observation}&{Seeing}&{Exposure}&\multicolumn{2}{c}{9\arcsec\ aperture}&\multicolumn{2}{c}{4\arcsec\ aperture}\\
{}&{}& &{Dates}&{(arcsec)}&{Time [s]}&{Mag.}&{Error}&{Mag.}&{Error}\\\hline
6C0825+34 &1.467 & J & 05/03/01                     	&0.7           & 1080 (1080) & 19.27 & 0.27& 19.86 & 0.17\\
          && H & 07/03/01, 16/02/02$^S$       	        &1.2, 0.7      & 1080 (540)  & 18.99 & 0.48& 19.68 & 0.29\\\vspace{-8pt}
          && K & 04/03/01, 06/03/01, 07/03/01 	        &1.3, 1.1, 1.1 & 9240 (4200) & 18.90 & 0.24& 19.12 & 0.12\\\\
6C0943+39 &1.035& J & 04/03/01                     	&0.9           & 2700 (2700) & 19.14 & 0.12& 19.70 & 0.12\\
          && H & 06/03/01 			 	&0.8           & 1080 (1080) & 18.64 & 0.17& 19.27 & 0.14\\\vspace{-8pt} 
          && K & 04/03/01, 05/03/01, 06/03/01 	        &0.9, 0.8, 1.0 & 5760 (3660) & 18.03 & 0.11& 18.09 & 0.07\\\\          
6C1011+36 &1.042& J & 05/03/01 			 	&0.7           & 1080 (1080) & 19.59 & 0.28& 19.68 & 0.20\\            
          && H & 06/03/01 			 	&0.8           &  540 (540)  & 18.60 & 0.24& 18.63 & 0.10\\\vspace{-8pt}
          && K & 05/03/01, 06/03/01 		 	&0.6, 0.9      & 4320 (3900) & 17.67 & 0.09& 17.83 & 0.06\\\\          
6C1017+37 &1.053& J & 05/03/01 			 	&0.6           &  960 (960)  & 19.35 & 0.20& 19.89 & 0.16\\            
          && H & 06/03/01, 06/02/02$^S$ 	 	&0.8, 0.6      & 3820 (3820) & 18.98 & 0.19& 19.54 & 0.15\\\vspace{-8pt}
          && K & 05/03/01, 06/03/01, 07/03/01 	        &0.6, 0.7, 1.2 & 5700 (3480) & 18.31 & 0.08& 18.57 & 0.09\\\\          
6C1019+39 &0.922& J & 04/03/01 			 	&0.8           & 1620 (1620) & 18.00 & 0.06& 18.47 & 0.07\\            
          && H & 06/03/01 			 	&0.8           &  480 (480)  & 17.37 & 0.09& 17.71 & 0.06\\\vspace{-8pt}
          && K & 04/03/01, 06/03/01 		 	&0.7, 0.7      & 4140 (2160) & 16.41 & 0.04& 16.80 & 0.04\\\\          
6C1100+35 &1.440& J & 05/03/01, 25/04/02$^B$ 	 	&0.7, 0.7      & 1620 (1620) & 19.45 & 0.19& 19.59 & 0.12\\            
          && H & 06/03/01 			 	&0.7           & 1080 (1080) & 18.70 & 0.19& 19.02 & 0.11\\\vspace{-8pt}
          && K & 05/03/01, 06/03/01, 26/04/02$^B$ 	&0.8, 0.7, 0.4 & 6960 (5340) & 18.19 & 0.12& 17.99 & 0.07\\\\          
6C1129+37 &1.060& J & 05/03/01, 25/04/02$^B$ 		&0.9, 0.7      & 2160 (2160) & 19.38 & 0.20& 19.35 & 0.11\\            
          && H & 06/03/01 				&0.7           &  480 (480)  & 18.24 & 0.23& 18.31 & 0.09\\\vspace{-8pt}
          && K & 04/03/01, 06/03/01, 24/04/02$^B$ 	&1.0, 0.7, 0.8 & 6240 (1620) & 17.63 & 0.11& 17.81 & 0.07\\\\          
6C1204+35 &1.376& J & 05/03/01, 25/04/02$^B$ 		&0.9, 0.7      & 2160 (2160) & 18.73 & 0.10& 19.31 & 0.11\\            
          && H & 14/04/01$^{S,I}$ 			&0.6           &  720 (720)  & 18.59 & 0.26& 18.76 & 0.11\\\vspace{-8pt}
  & &K & 05/03/01, 06/03/01, 07/03/01, 22/04/02$^B$&1.0, 0.6, 0.9, 0.4 & 7500 (6900) & 17.85 & 0.08& 18.01 & 0.07\\\\          
6C1217+36 &1.088& J & 04/03/01 				&1.2           & 1560 (1560) & 19.26 & 0.16& 19.50 & 0.12\\            
          && H & 14/04/01$^{S,I}$ 		        &0.6           & 1440 (1440) & 17.95 & 0.13& 18.51 & 0.07\\\vspace{-8pt}
          && K & 04/03/01, 06/03/01, 07/03/01 	        &0.8, 0.6, 0.9 & 6960 (1140) & 17.22 & 0.08& 17.55 & 0.06\\\\          
6C1256+36 &1.128& J & 04/03/01 				&1.2           & 1440 (1440) & 19.11 & 0.16& 19.74 & 0.20\\            
          && H & 07/03/01 				&0.6           &  540 (540)  & 18.11 & 0.14& 18.58 & 0.09\\\vspace{-8pt}
          && K & 04/03/01, 06/03/01, 07/03/01           &0.8, 0.8, 0.7 & 7800 (4620) & 17.45 & 0.07& 17.83 & 0.06\\\\          
6C1257+36 &1.004& J & 05/03/01 				&1.0           & 1080 (1080) & 19.15 & 0.17& 19.39 & 0.11\\            
          && H & 14/04/01$^{S,I}$ 		        &0.6           &  720 (720)  & 18.04 & 0.17& 18.27 & 0.07\\\vspace{-8pt}
          && K & 05/03/01, 06/03/01, 07/03/01 	        &1.0, 0.7, 0.6 & 7260 (5760) & 17.17 & 0.05& 17.50 & 0.05
\end{tabular}		            
\end{center}		            
\end{table*}		           
\normalsize
 
All observations used a nine point jitter pattern, with offsets of
roughly 10\arcsec\ between each 1 minute exposure (with the exception
of the IRCAM3 service observations, which used only 4 pointings per
jitter pattern).  The observational data were dark subtracted, and
masked for bad pixels.  Blocks of up to 120 frames of data were
combined and median filtered to create a first-pass sky flat-field
image, which accounted for the
majority of the pixel-to-pixel variations of the chip.  However,
large-scale illumination gradients remained, due to the changing position of
the telescope over the night.  To correct for this, smaller groups of
9 -- 18 first pass flat-fielded images were similarly combined to create
residual sky flat-field images, which successfully 
accounted for this effect.  Bright objects on the flat-fielded images
were then masked out, and the process repeated, allowing the data
to be cleanly flat fielded without any contamination from stars or
galaxies. 
 
The flat-fielded data for each source were sky-subtracted and combined
using the IRAF package DIMSUM, creating a final mosaiced image of
approximately $115 \times 115$ arcsec$^2$\footnote{The highest
signal-to-noise level is restricted to the central $70 \times 70$
arcsec$^2$.}, which was flux calibrated using observations of UKIRT faint
standard stars.  Photometry was 
carried out within 9\arcsec\ and 4\arcsec\ diameter apertures using
the IRAF package \textsc{apphot} and a single sky annulus, correcting  
for the presence of companion objects in the larger 9\arcsec\ apertures only.
Galactic  extinction corrections and uncertainties for the resulting
magnitudes were determined in the same way as for the HST observations.
The resulting magnitudes presented in Table 2 are generally in very
good agreement with previous $K$ band photometry of these galaxies
(Lilly, Longair \& Allington-Smith 1985; Eales et al 1997).

\subsection{Astrometry}

Astrometric solutions for the pixel scales and rotations of the HST
and UKIRT images were determined using the method outlined below.
Firstly, the pixel position of unsaturated stars present in both the
image and in the APM database (Maddox {\it et al} 1990) were
registered.  As the 6C sources were selected at a significant distance
from the plane of the Milky Way, a sufficient number of APM stars were
not always available in the field of view of the observations.  Where
an inadequate number of unsaturated stars were available, compact
elliptical galaxies with APM positions were also used if practical.
The data were used to calculate an accurate measure of the pixel
scales and the angle of rotation of the coordinate system. 
 
This first method worked fairly well for the majority of the sources.
For a few sources however, no reasonable fits were achievable due to a
lack of sufficient APM stars.  In these cases, the UKIRT pixel scale
and rotation were fixed at the mean of the values determined in other
observations (these are stable at $< 1^\circ$) and the position of
bright objects measured from the digitised sky survey was used to fit
the right ascension and declination of the image.  Comparison of the
pixel positions of the numerous APM unidentified objects common to
both the UKIRT and HST images then allows accurate determination of
the HST astrometry. 
 
Using both these methods where necessary, the astrometric errors are
estimated to be less than about 0.5$^{\prime\prime}$, and usually better
than 0.25$^{\prime\prime}$.  The HST and UKIRT
images of reach source were spatially aligned to an accuracy of better
than 0.1$^{\prime\prime}$. 

\begin{table*}
\caption{Contributions of line emission and continuum emission to the
HST magnitudes. The fifth column lists the emission lines which lie
within the spectral range of the observations, and the sixth the
percentage of the total magnitude which can be accounted for by line
emission, plus the associated error.  The seventh column gives the percentage of the
total flux in each filter which can be accounted for by nebular
continuum emission and the associated error.  Where only an upper
limit is available for the $H\beta$ flux, the nebular continuum
percentage is denoted as an upper limit. }
\begin{center} 
\begin{tabular} {lcccccrlrlc}
{[1]}&{[2]}&{[3]}&{[4]}&{[5]}&&\multicolumn{2}{c}{[6]} &\multicolumn{2}{c}{[7]}&\\
{Source}&{Filter}&{9\arcsec\
WFPC}&Error&{Emission}&&\multicolumn{2}{r}{Line Emission}
&\multicolumn{2}{r}{Nebular Continuum}&\\
        &        &{Mag.}&     &{Lines}   &&{Flux}&{Error}     &{Flux}&{Error}& \\\hline

6C0825+34 & F814W & 21.65 & 0.15 &\footnotesize{[NeV],[OII],[NeIII],H$\zeta$,[NeIII]+H$\epsilon$}    & &  $3\%$ & $\pm2\%$&   $<16\%$&$\pm3\%$&\\
6C0943+39 & F702W & 21.78 & 0.38 &\footnotesize{[NeV],[OII],[NeIII],[NeIII]+H$\epsilon$  }     &  & $18\% $  &$\pm5\%$ & $<6\%$&$\pm^{1\%}_{0\%}$&\\
6C1011+36 & F702W & 21.59 & 0.19 &\footnotesize{[NeV],[OII],[NeIII],H$\zeta$,[NeIII]+H$\epsilon$,H$\delta$ }   & & $8\% $ & $\pm4\%$ & $16\%$&$\pm5\%$&\\
6C1017+37 & F702W & 21.53 & 0.16 &\footnotesize{[NeV],[OII],[NeIII],H$\zeta$,[NeIII]+H$\epsilon$,H$\delta$} & &  $17\%$ & $\pm4\%$&   $25\%$&$\pm6\%$&\\
6C1019+39 & F606W & 21.89 & 0.25 &\footnotesize{[NeV],[OII] }          &&$3\%$&$\pm2\%$&   $<3\%$&--&\\
          & F814W & 19.86 & 0.10 &\footnotesize{[OII],[NeIII],H$\beta$}&&$5\%$&$\pm1\%$& $<0.5\%$&--&\\\vspace{1.5pt}
6C1100+35 & F814W & 21.80 & 0.18 &\footnotesize{[NeV],[OII],[NeIII],H$\zeta$,[NeIII]+H$\epsilon$,H$\delta$} & &   $7\%$  & $\pm4\%$ &  $<1\%$&$\pm^{0\%}_{0.5\%}$&\\
6C1129+37 & F702W & 21.52 & 0.14 &\footnotesize{[NeV],[OII],[NeIII],[NeIII]+H$\epsilon$                     } & &  $15\%$  &$\pm5\%$ &   $<22\%$&$\pm^{3\%}_{0\%}$&\\
6C1204+35 & F814W & 20.79 & 0.29 &\footnotesize{[NeV],[OII],[NeIII],H$\zeta$,[NeIII]+H$\epsilon$,H$\delta$  } &  & $11\%$  & $\pm7\%$  & $<31\%$&$\pm^{4\%}_{2\%}$&\\
6C1217+36 & F606W & 22.26 & 0.32 &\footnotesize{CII],[NeIV],[NeV]
}     & &   $0\%$  &  $+1\%$&  $<1\%$&--&\\
 & F814W          & 20.59  & 0.12 &\footnotesize{[NeV],[OII],[NeIII],[NeIII]+H$\epsilon$,H$\gamma$        }    & &   $3 \%$&$\pm1\%$ &   $<1\%$&--&\\\vspace{1.5pt}
6C1256+36 & F702W & 22.23 & 0.16 &\footnotesize{[NeV],[OII],[NeIII],H$\zeta$                              }   &  & $9\%$  &$\pm4\%$ &   $33\%$&$\pm^{10\%}_{9\%}$&\\
6C1257+36 & F606W & 22.04 & 0.26 &\footnotesize{CII],[NeIV],[NeV]                                          }  & &   $2\%$ &  $\pm1\%$ & $27\%$&$\pm^{10\%}_{9\%}$&\\
 & F814W          & 20.55  & 0.12 &\footnotesize{[OII],[NeIII],H$\zeta$,[NeIII]+H$\epsilon$,H$\gamma$,H$\beta$}&  & $10\%$ & $\pm2\%$  & $25\%$&$\pm^{13\%}_{8\%}$&
 \end{tabular}		           
\end{center}		           
\end{table*}		           

\section{Contribution of line emission and nebular continuum emission
to the HST magnitudes}
\subsection{Percentage contribution of line emission to the HST magnitudes}

Due to the fact that for each source, several prominent emission lines
lie within the wavelength range of the HST filters, it is important to
estimate the contribution of line emission to the HST magnitudes.
The emission line contribution was derived from spectroscopic
observations of each source, assuming that the distributions of line
and continuum emission were comparable, and so the equivalent widths of the
emission lines in the spectral slit region were comparable to those in
the entire source. For the eight sources with $z < 1.2$ the deep
spectroscopic observations of Inskip et al (2002a) were used. For the
remaining three higher redshift sources, the shallower Rawlings et al (2001)
spectra were used.   The results of this analysis are tabulated in Table 3.  

The line emission contribution to the total UV excess varies from
3-18\%, and as expected is less on
average for the 6C sources than for the more powerful 3CR radio sources observed by Best
et al (1997) at the same redshift.  However, due to
the lower radio power of the 6C sample, and the evidence that emission
line flux scales with radio power (e.g. Inskip {\it et al} 2002a,
Jarvis {\it et al} 2001), one might expect line emission to account
for an even lower fraction of the total optical/UV excess observed in the 6C
observations than is actually observed.  However, the choice of filters is such that the
wavelength of the dominant \oo\ emission line is at a higher filter
throughput for the 6C observations than the 3CR observations. 
As this is the most prominent emission
line in the spectra of these galaxies at the wavelengths in question, the emission
line contribution initially appears to be higher than might be
expected for the 6C sources.   

\subsection{Spatial variation of emission line contribution}

As fairly large emission line contributions are measured for several
sources, it is well worth estimating how these emission lines
contribute to the extended aligned emission.  This
can only be effectively done for the sources with emission line
regions extended on scales significantly larger than the seeing, for
which good 2-d spectra are available, i.e. 6C0943+39 and 6C1129+37.
Although 6C1017+37 also has a strong emission line contribution, this
source is not considered as it is not particularly extended in either
the spectra or in the HST image. 
 
For both sources, the HST image was convolved with a Gaussian in order
to approximate the 1.5\arcsec\ ground-based seeing conditions of the
spectra.   Then, the amount of flux in the HST image which would fall
within the 2\arcsec\ wide slit used for our spectroscopic observations
was obtained.  This was separated into three regions: a central
4\arcsec\ region centred on the radio galaxy, and two regions either
side, from 2\arcsec\ out to 4.5\arcsec\ from the centre of the galaxy
(this corresponds to the limiting extent of the 9\arcsec\ diameter
apertures used in the photometry).   Next, the 2-d surface brightness
profile of the \oo\, emission line (Inskip et al 2002a) was used to
determine the emission line flux for the central $4^{\prime\prime}$,
and the regions outside.  These values were used to determine how the
ratio of emission line flux to total flux varied in the central, east
and west regions.
 
For 6C0943+39, the HST flux falling in the central region of the slit
was found to be 17.5\% line emission.  The outer regions were found to
be similar on either side of the galaxy; here emission lines account
for 39.5\% of the total flux (c.f. 18\% line emission in total). 88\%
of the total emission lies within the central region, whereas 91\% of
the total continuum emission and 77\% of the total line emission lie
within this region.  For this source, the extended faint emission
includes a stronger contribution from line emission. 
 
For 6C1129+37, the HST flux falling in the central region of the slit
was found to be 20\% line emission. The flux in the 2.5\arcsec\ region
of the slit to the west of the host galaxy was found to be 7.8\% line
emission, the flux in the east region is 13.4\% line emission
(c.f. 15\% line emission in total). 
Clearly, for this source, emission lines do not
dominate the more extended regions of emission.  

\subsection{Nebular continuum emission}

Emission lines clearly provide a significant fraction of the flux
observed in the HST images.  In addition to line emission, nebular
continuum radiation is also produced due to other radiative processes 
associated with the ionized gas. (These include free--bound
(recombination) and free--free (bremsstrahlung) transitions involving
hydrogen and helium, and the two--photon decay of the 2 $^{2}S$ level
of hydrogen.) It is likely that nebular
continuum emission will be an important factor for some sources.
The level of the nebular continuum emission can be defined as a function of 
wavelength (in Angstroms) as:\\
$F_{\nu}(neb) = \frac{3\times10^{-22} F(H\beta) \gamma_{tot}(\lambda)}{(1+z) \lambda^2 B}$\\
where $\gamma_{tot}(\lambda)$ is the total emission coefficient for
the different physical processes which cause the nebular continuum
emission, $F(H\beta)$ is the $H\beta$ flux and $B$ is a constant
($1.24\times10^{25}$ for an assumed temperature of $10^4$K).  Values
for these parameters have been taken from Aller {\it et al}
(1987). This method for calculating the nebular 
continuum level is also the same as that used by Dickson {\it et al} (1995).
$H\beta$ fluxes are the same as those used by Inskip et al (2002b) for the
sources at $z < 1.2$. For the higher redshift 6C sources an upper
limit on $H\beta$ has
been calculated from the average line ratios tabulated in Inskip et al
(2002a) and the emission line fluxes of Rawlings {\it et al} (2001).
 
The contribution of the nebular continuum to the total flux in each
filter has been calculated using the formula above, and the results
listed in Table 3. The derived percentages typically vary by $< 3\%$
if alternative temperatures of 5000 or 20000K are assumed instead;
this has been added in quadrature with the $H\beta$ flux uncertainty
to give the error values in column 7. 
For the sample as a whole, the nebular continuum contributions range
from 0\% to 33\% of the total flux.  This range is comparable to the
3\% to 40\% contribution of nebular continuum emission to the total UV
continuum flux found by Tadhunter {\it et al} (2002) for a complete, unbiased
sample of $0.15 < z < 0.7$ 2-Jy radio galaxies.
 
For 6C0943+39 and 6C1129+37, the nebular continuum contribution can
also be estimated as a function of position. The percentage
contribution of nebular continuum emission is 6\% for the central
regions of 6C0943+39, rising to 13\% outside the central 4\arcsec.
For 6C1129+37, nebular continuum emission contributes 29\% of the
total flux in the central region, 11\% in the west region and 20\% in
the east region.  

\begin{figure*}
\vspace{5.55 in}
\begin{center}
\includegraphics{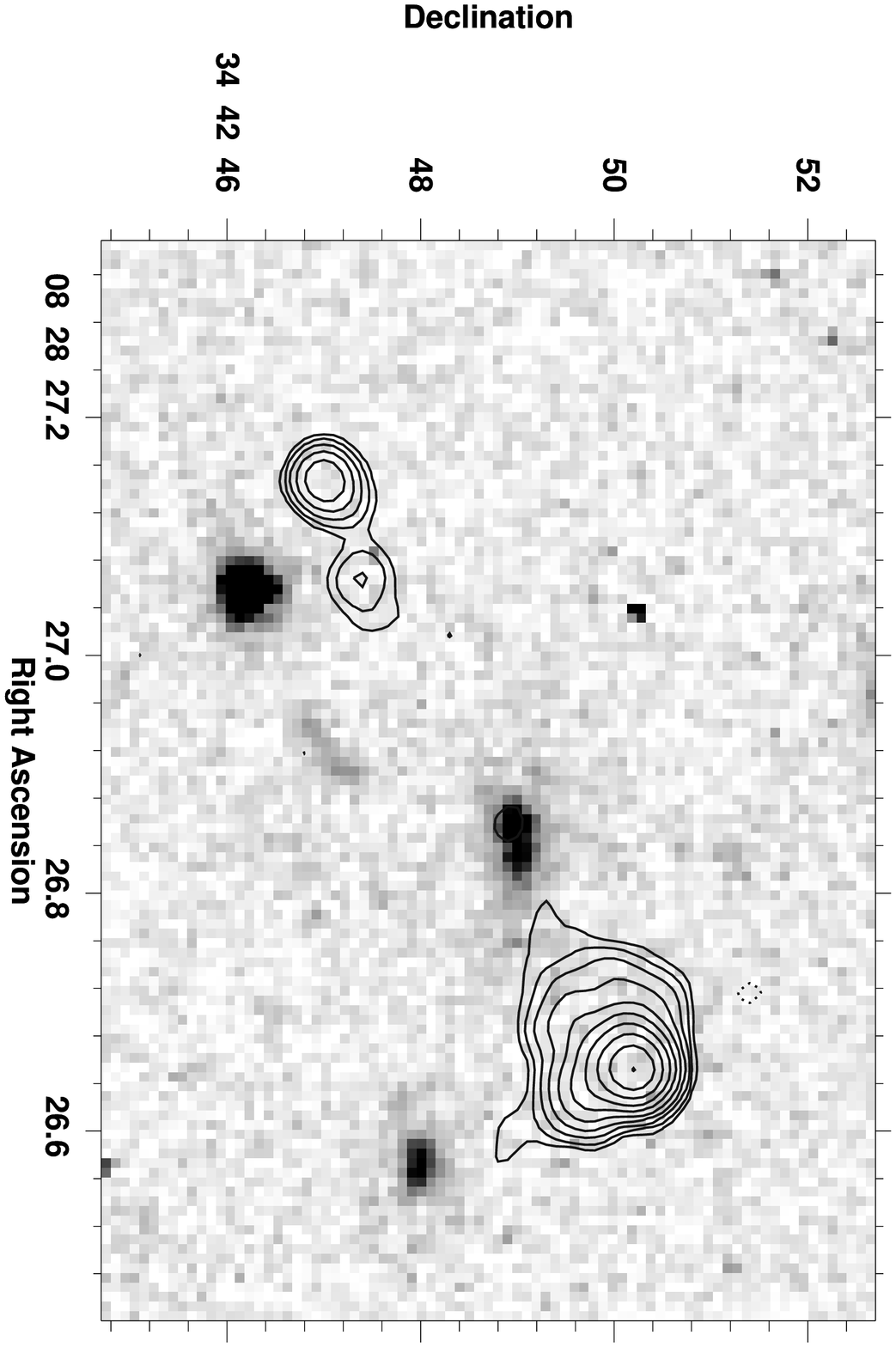}
\includegraphics{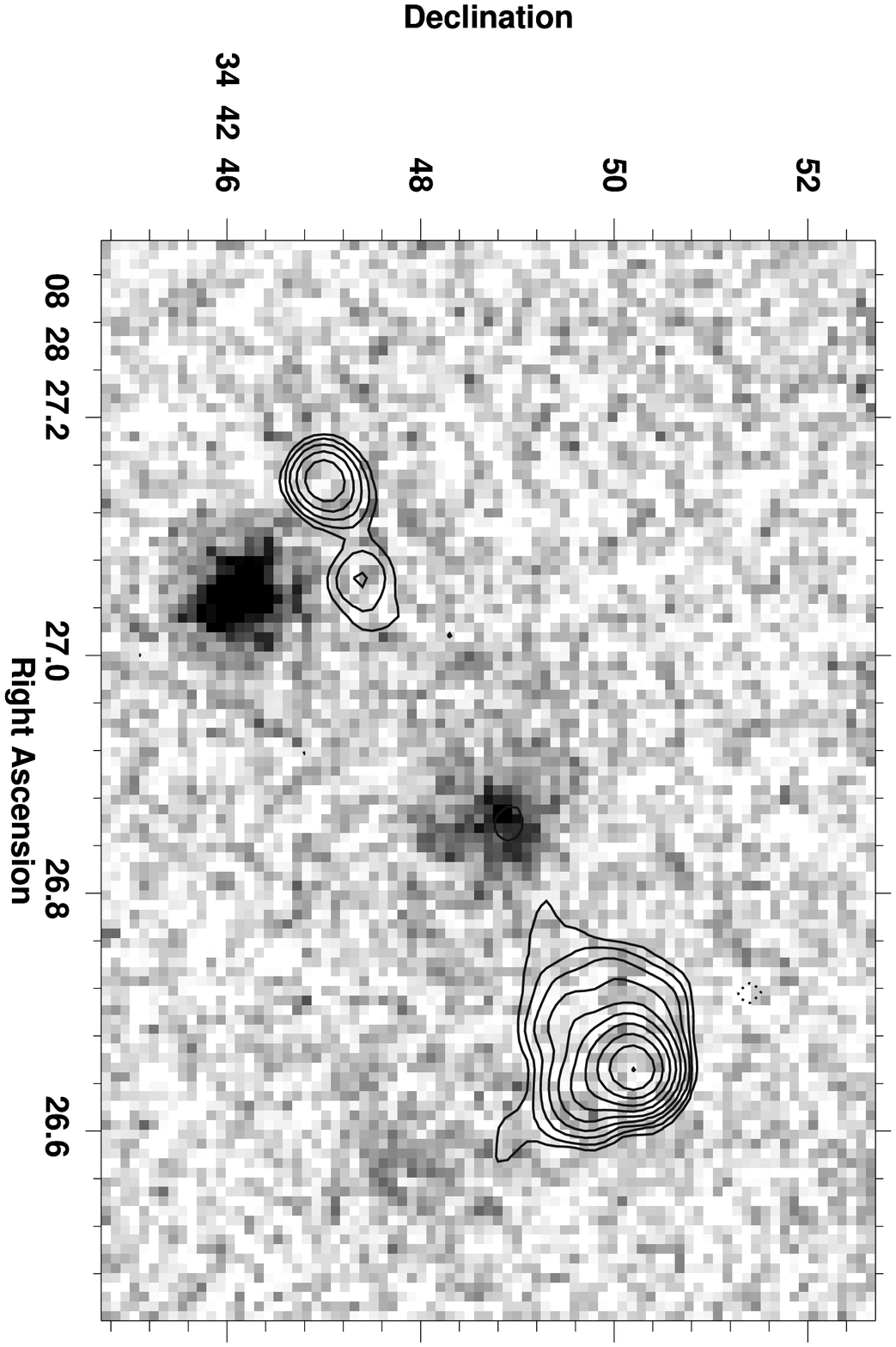}
\end{center}
\caption{6C0825+34: (a - top) $K$-band image of 6C0825+34. (b - bottom)
HST F814W image of 6C0825+34.  Contour lines for both images represent the
5GHz VLA observation of this source, with contours at 100$\mu\rm{Jy}$ beam$^{-1} \times$ (-1,1,2,4...1024).  All coordinates are in epoch J2000.0.
\label{Fig: 6_1}}
\end{figure*}
\section{Multi-wavelength images of the 6C \small{\textbf{z}} \normalsize{}\textbf{$\sim 1$ subsample}}

In this section the images of the galaxies in the optical, infrared
and radio wavebands are presented (Figs.~\ref{Fig: 6_1}--\ref{Fig:
6_11}).  The HST and UKIRT $K$-band images have been overlaid 
with 5GHz Very Large Array (VLA) radio contours.     
  Zoomed-in images of the host galaxy are also provided
for the larger radio sources in the sample. 
For the two sources which seem to have very close companion galaxies
(6C1129+37 and 6C1256+36), close-ups of the HST images
with UKIRT $K$-band contours are also presented. 
 
The eleven sources are presented in order of increasing right ascension.
All coordinates are in equinox J2000.0.  A brief description of the key
features of each source is also provided.

\subsection*{6C0825+34}

6C0825+34, at $z = 1.46$, is the faintest and most distant galaxy in
the sample.  The UKIRT $K$-band image (Fig.~\ref{Fig: 6_1}) shows an
approximately elliptical 
host galaxy. A faint tail of aligned emission can be seen on the HST 
F814W image.   6C0825+34 has
much bluer $J-K$ and $H-K$ colours than the other galaxies in the
sample.  Its F814W$-K$ colour is average for the 6C sample, but low
when compared to other 6C or 3CR sources at similar redshifts.  The
very blue colour of this source could be explained by  some
contamination from an incompletely obscured AGN.   The deep VLA radio
observations made of this source are also interesting (Best {\it et
al} 1999) in terms of interpreting the observed colours.  The north
west radio lobe displays little depolarization 
at 5GHz, whereas the south east lobe is completely depolarized.
Whilst this
strong asymmetry in the flux levels of the radio lobes would 
favour a radio axis at a relatively large angle to the plane of the
sky, the lack of a strong radio core argues against this
interpretation.   Additionally, the radial profile of the HST emission
(see Paper 2 for the full analysis of the radial profiles)
shows no evidence for a point source component, suggesting that a
different interpretation is required for the very blue F814W$-K$ colour of
this source.   
\begin{figure*}
\vspace{4.75 in}
\begin{center}
\includegraphics{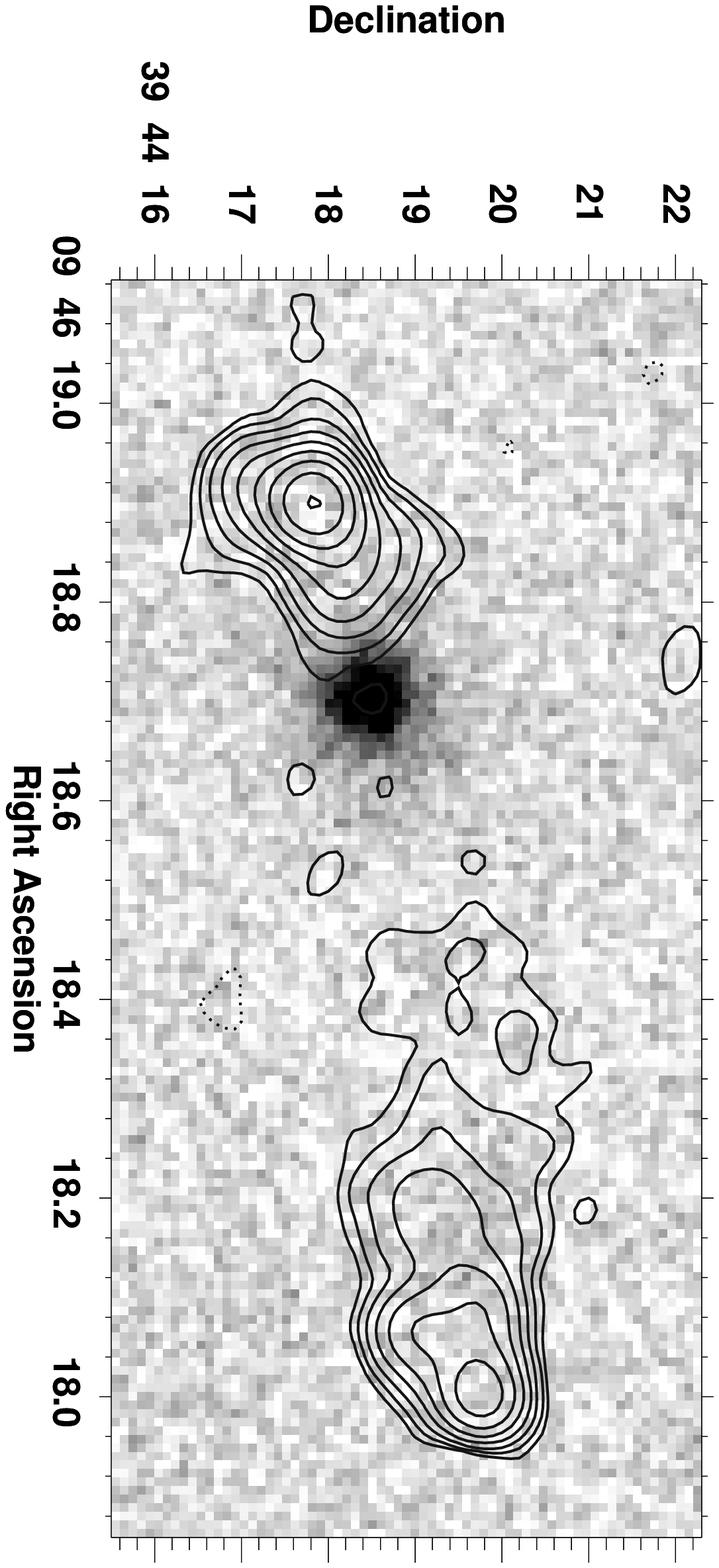}
\includegraphics{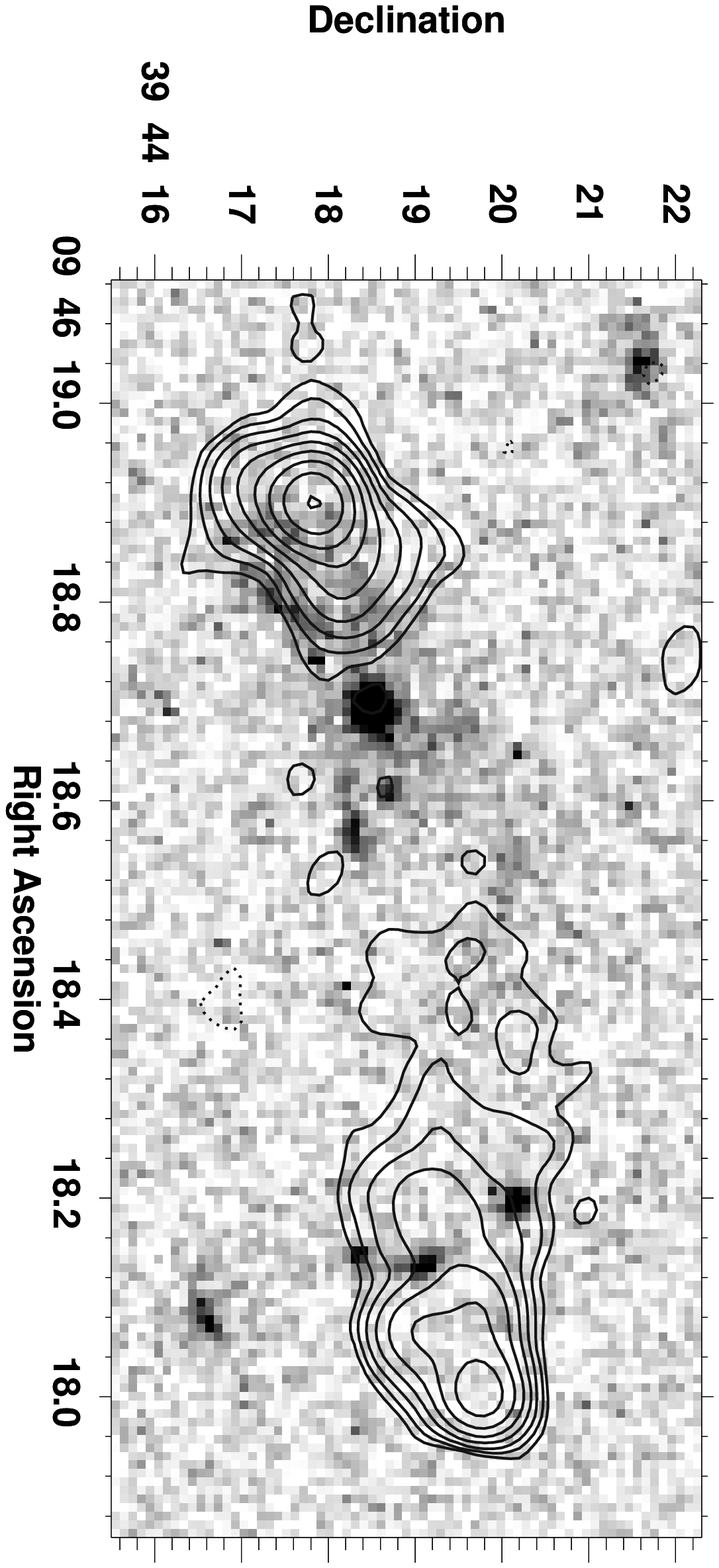}
\end{center}
\caption{6C0943+39: (a - top) $K$-band image of 6C0943+39. (b - bottom)
HST F702W image of 6C0943+39.  Contour lines for both images represent the
5GHz VLA observation of this source, with contours at 90$\mu\rm{Jy}$ beam$^{-1} \times$ (-1,1,2,4...1024).  All coordinates are in epoch J2000.0.
\label{Fig: 6_2}}
\end{figure*}

\subsection*{6C0943+39}

6C0943+39, at $z = 1.036$, is clearly shown to be a bright elliptical
galaxy in the $K$-band (Fig.~\ref{Fig: 6_2}).  The colours of this
source are average for the 
sample. The HST image shows extensive faint aligned emission
surrounding the elliptical host galaxy.   The more extended features 
include a strong contribution from line emission and nebular continuum
emission ($\sim 50\%$ in total; see
Section 3).  At a distance of roughly 5$^{\prime\prime}$ to the east
of the elliptical host galaxy (outside the 9$^{\prime\prime}$ diameter
aperture used for the photometry), three bright emission regions can
be seen, which may possibly be associated with the radio source. These
appear to lie within the radio lobe, and form a line roughly 
perpendicular to the radio axis. Line emission is also observed in
this region, with a narrower line width than is seen closer to the
host galaxy. The colours of these three features
are relatively blue in comparison with the radio galaxy
itself.

With a projected angular size of 90kpc, 6C0943+39
is one of the smaller radio sources in the subsample.  A faint radio
core is observed at 5GHz, coincident with the elliptical host galaxy.
The radio observations (Best et al 1999) also show that the eastern lobe is strongly
depolarized, and that there may be a sharp bend in the radio emission
near the hotspot.  
 
The spectroscopic observations of this source are consistent with
shocks as the major ionization mechanism, which may perhaps account
for the greater importance of line emission in the extended features
surrounding this galaxy (Inskip et al 2002a).  The odd L-shaped two
dimensional profile of the \oo3727 emission line for this source may
reflect the morphology of the aligned structures seen in the HST
observations: the broader line emission may be associated with the
diffuse UV emission aligned with the radio axis in the central
4$^{\prime\prime}$ region, and the less extreme gas kinematics may
perhaps be linked to the bright knotty features to the west of the
host galaxy.   

\subsection*{6C1011+36}

\begin{figure*}
\vspace{6.6 in}
\begin{center}
\includegraphics{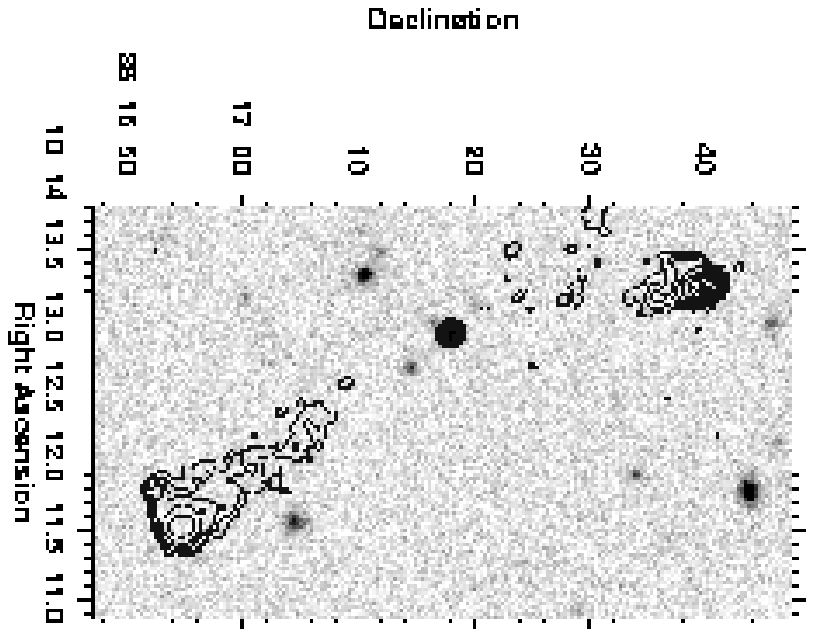}
\includegraphics{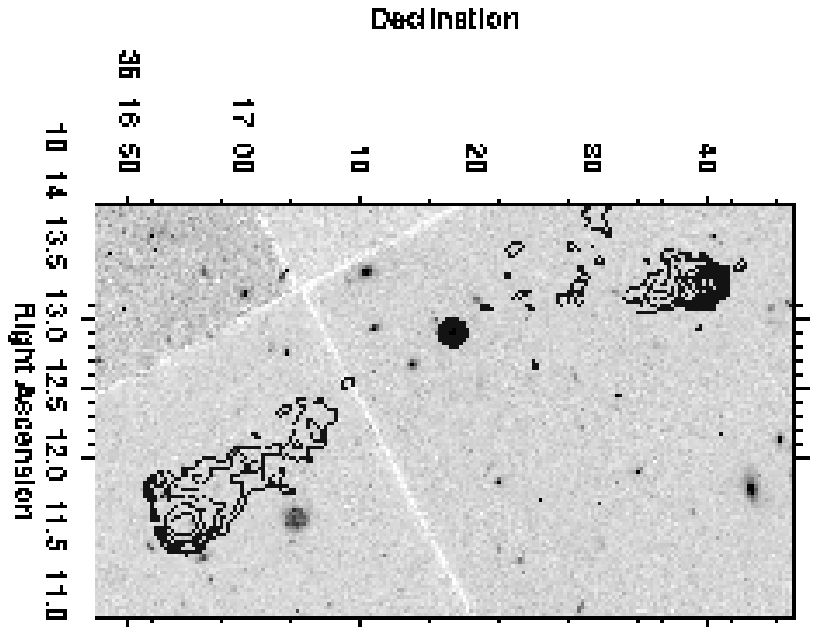}
\includegraphics{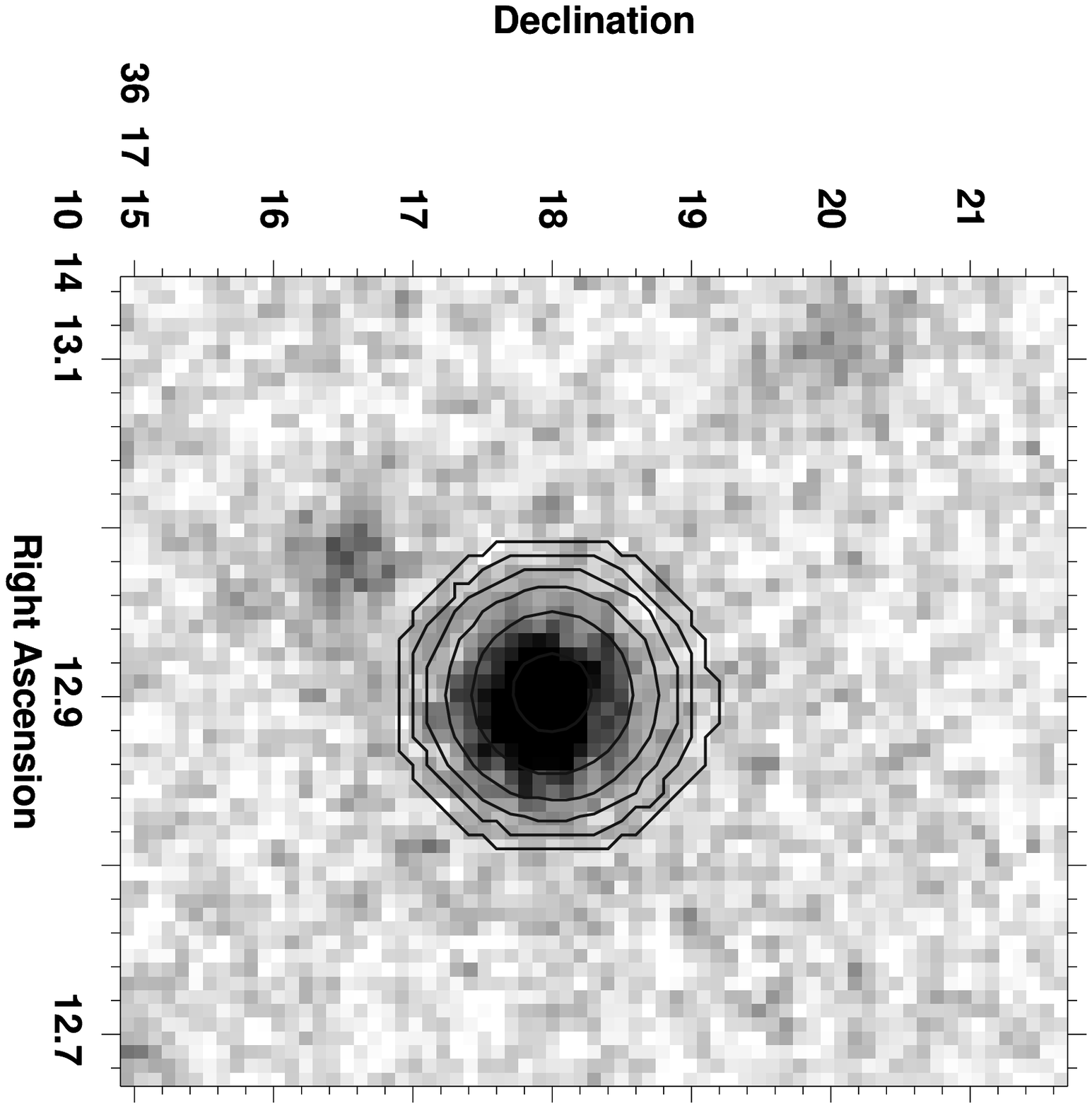}
\includegraphics{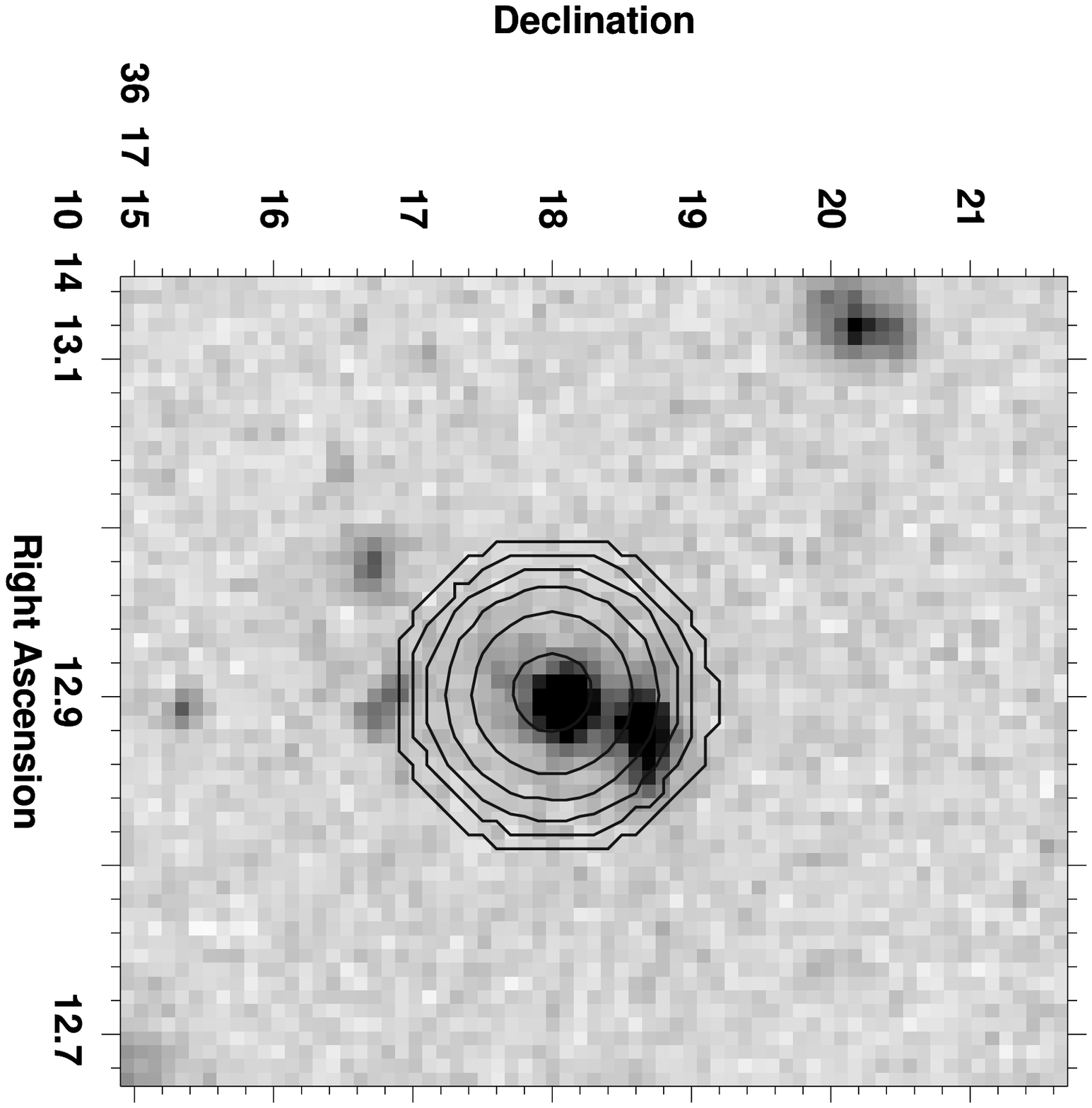}
\end{center}
\caption{6C1011+36: (a - top left) $K$-band image of 6C1011+36. (b - 
top right) HST F702W image of 6C1011+36. (c - bottom left) zoomed in
$K$-band image of the central region of 6C1011+36.  (d -
bottom right) zoomed in HST F702W image of the central region of
6C1011+36.  Contour lines for all images represent the 5GHz VLA
observation of this source, with contours at 80$\mu\rm{Jy}$ beam$^{-1} \times$ (-1,1,2,4...1024).  All coordinates are in epoch J2000.0. 
\label{Fig: 6_3}}
\end{figure*}

6C1011+36, at $z = 1.042$, is the largest radio source in the sample
with a projected size of 444kpc.  The relative contribution of the
core flux density to the total flux density is high, and a double hotspot is observed in the southern radio
lobe.  The HST image (Fig.~\ref{Fig: 6_3}) shows the elliptical host
galaxy and a very bright blue feature towards the north west, loosely
aligned with the radio source axis. This feature is not observed in
the $K$-band observations of 6C1011+36.  Due to its proximity to the
host galaxy it is difficult to
separate the flux due to this feature from that of the host galaxy 
itself. However, an analysis of the flux in a small 0.6$^{\prime\prime}$
diameter aperture centred on this feature gives a very blue colour of
F702W$-K \lta 1.7$. 
Additionally, there are several other
galaxies near this source, some of which are in very close proximity
to the host galaxy.   Roche, Eales \& Rawlings (1998) suggest that
this source is currently undergoing a merger with a much smaller
galaxy seen to the south east in the $K$-band image, and may be
interacting with its other nearby companions, which are also observed
in our imaging observations. Roche, Eales \& 
Hippelein (1998) have analysed the clustering properties of most of
the galaxies in our $z \sim 1$ 6C subsample.  The
cross--correlation amplitudes were larger than average for 6C1011+36,
consistent with Abell 2 clustering around this source, unlike the
Abell class 0 environments found for the majority of the sources in
this sample.  The line
emission from this source does not dominate the aligned emission, and
is well explained purely by photoionization by an obscured AGN.  
\begin{figure}
\vspace{5.9 in}
\begin{center}
\includegraphics{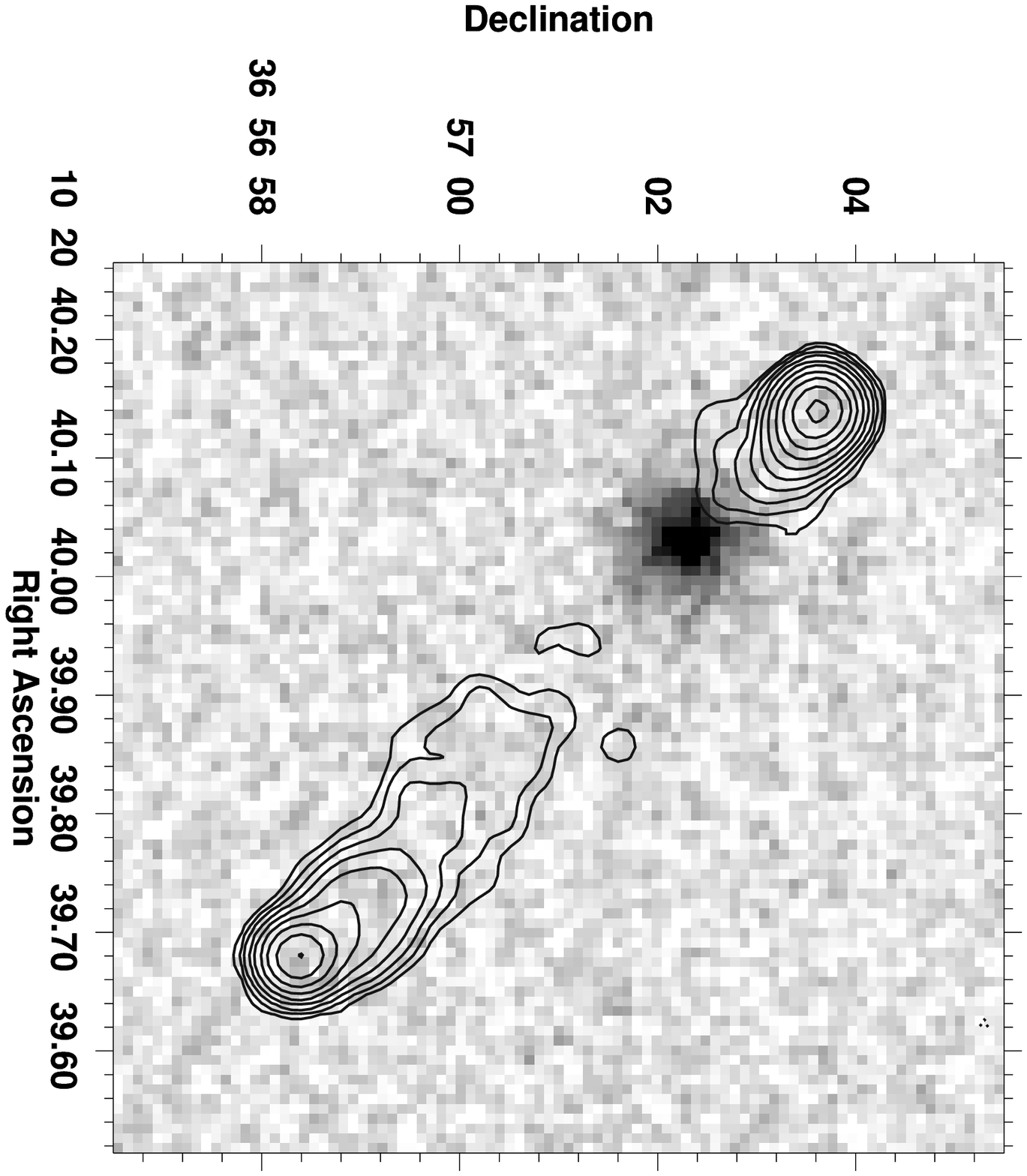}
\includegraphics{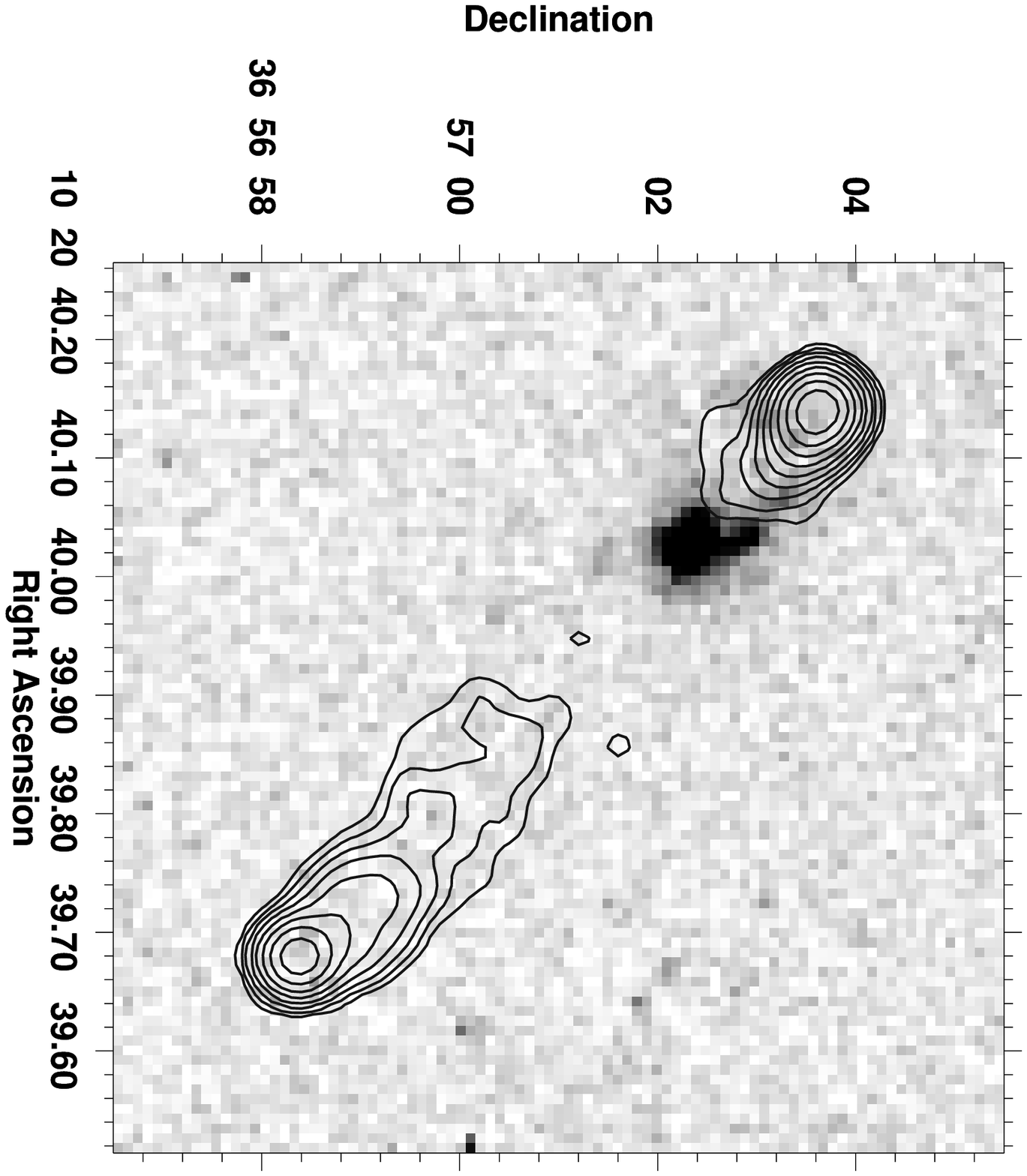}
\end{center}
\caption{6C1017+37: (a - top) $K$-band image of 6C1017+37. (b - bottom)
HST F702W image of 6C1017+37.  Contour lines for both images represent the
5GHz VLA observation of this source, with contours at 80$\mu\rm{Jy}$ beam$^{-1} \times$ (-1,1,2,4...1024).  All coordinates are in epoch J2000.0.
\label{Fig: 6_4}}
\end{figure}
\begin{figure}
\vspace{8.3 in}
\begin{center}
\includegraphics{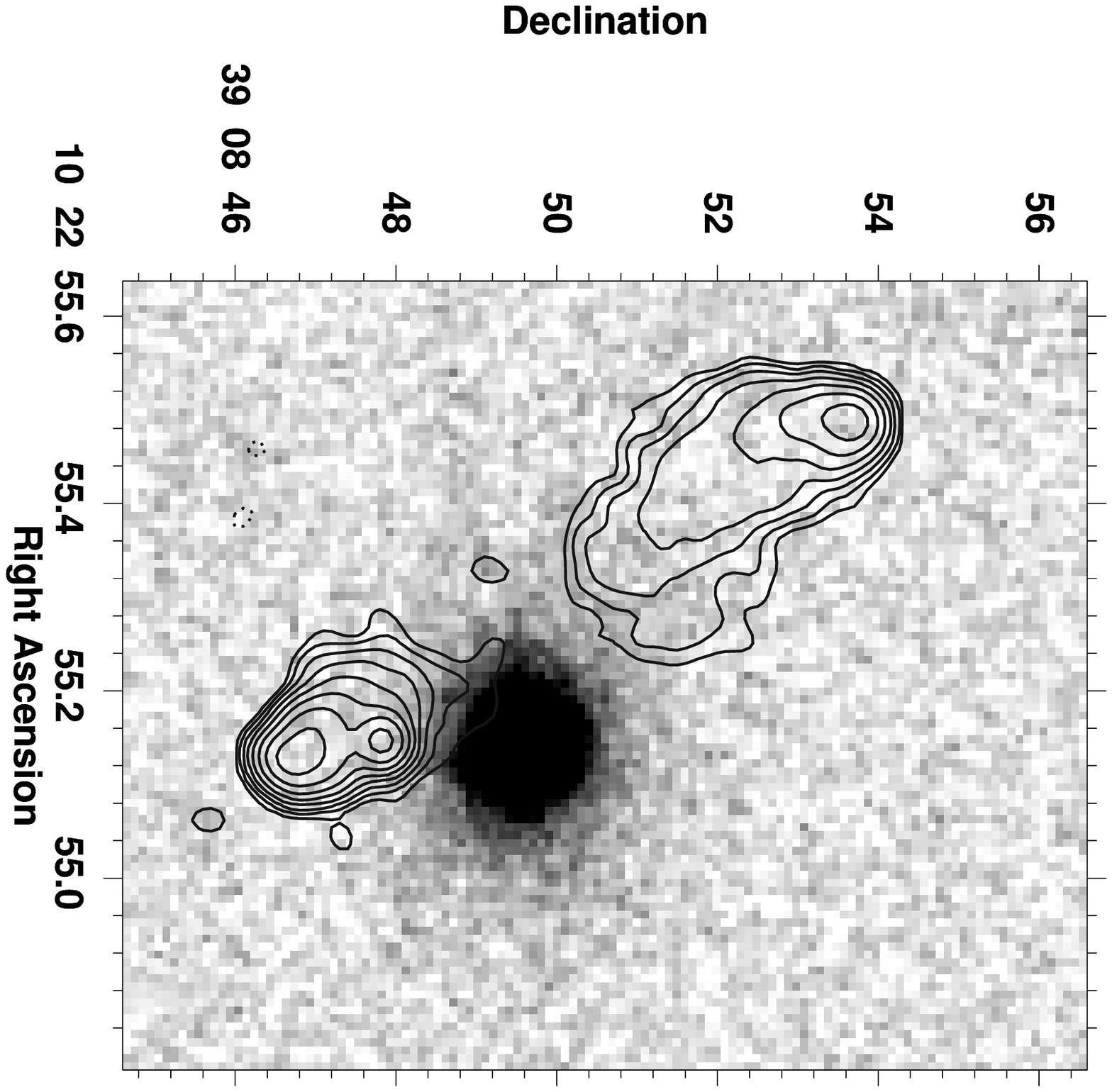}
\includegraphics{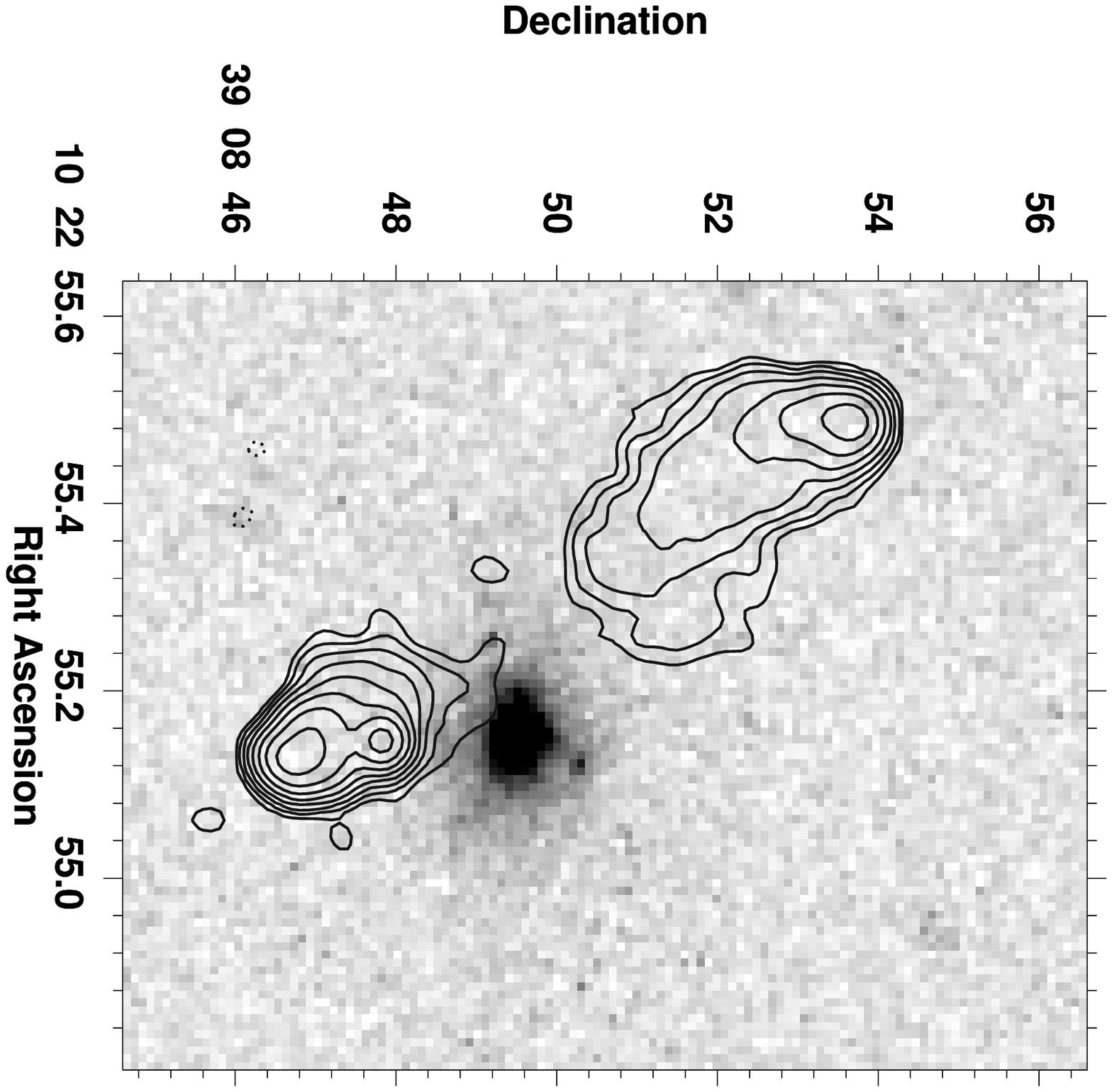}
\includegraphics{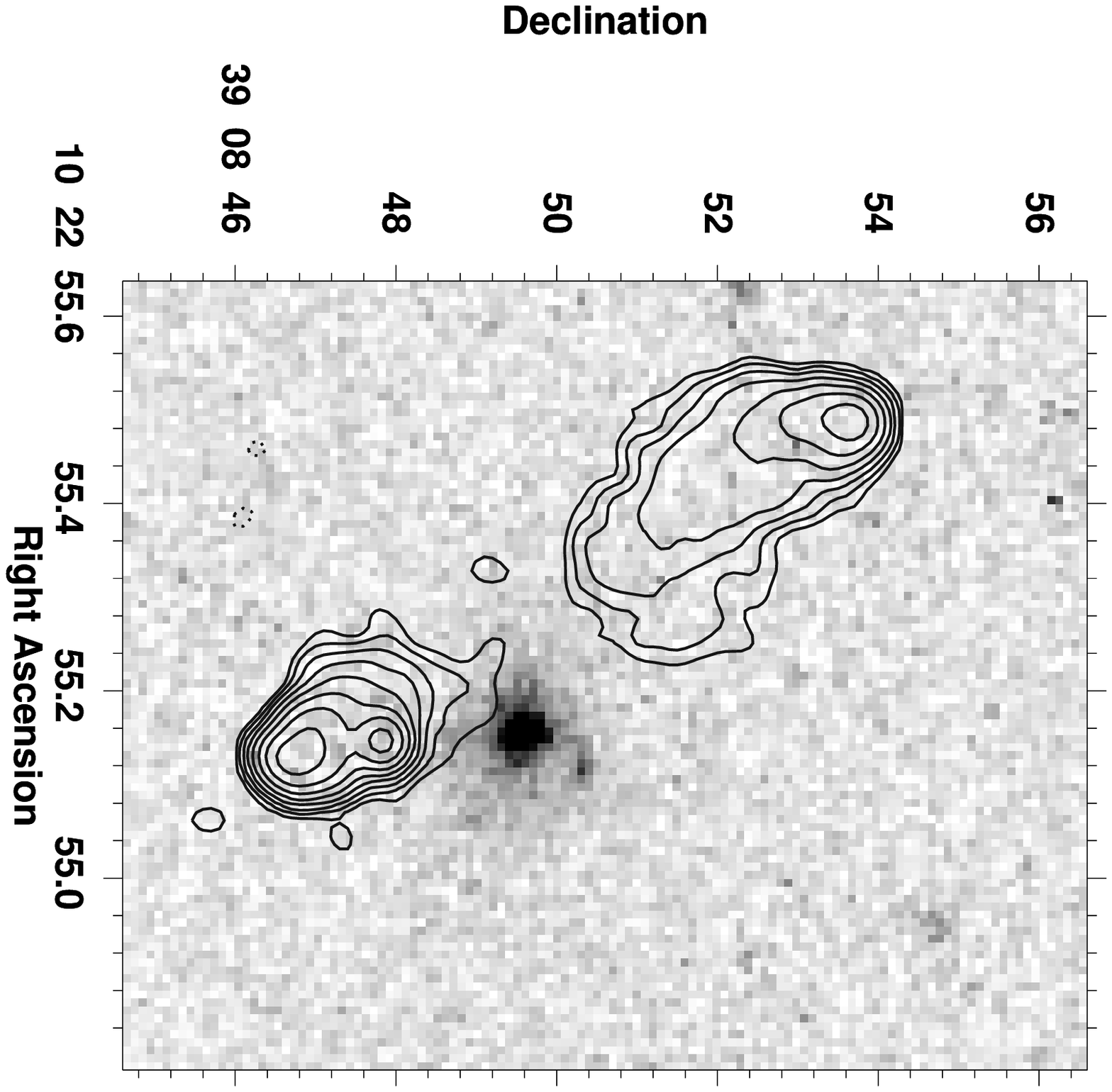}
\end{center}
\caption{6C1019+39: (a - top) $K$-band image of 6C1019+39. (b - 
centre) HST F814W image of 6C1019+39.  (c - bottom) HST F606W image of
6C1019+39.  Contour lines for all images represent the 5GHz VLA 
observation of this source, with contours at 90$\mu\rm{Jy}$ beam$^{-1} \times$ (-1,1,2,4...1024).  All coordinates are in epoch J2000.0. 
\label{Fig: 6_5}}
\end{figure}

\subsection*{6C1017+37}

6C1017+37, at $z = 1.053$, has an elliptical appearance in both the
HST and UKIRT images (Fig.~\ref{Fig: 6_4}).  Although no radio core is apparent at 5GHz, a
faint core was detected by Best {\it et al} (1999) in their 8GHz
observation.  This source is highly asymmetric in the angular sizes of
its radio lobes, and its rotation measures.   The HST image shows an
arc of aligned emission extending into the shorter, north--eastern
radio lobe.  The HST emission contains a strong contribution from line
emission and nebular continuum emission.  Spectroscopic observations
of this source show that its emission line region is fairly compact,
and displays a very large velocity width ($\sim 900\rm{km}\,\rm{s}^{-1}$).  The spectra are best
explained by a mixture of ionization mechanisms: both ionization by
radio source shocks and photoionization by an obscured AGN (Inskip et
al 2002a).  When compared with the data for other radio galaxies at
this redshift, the $J-K$ colour obtained from the UKIRT images for
this source is particularly blue. In addition to this, the F702W$-K$
colour for this galaxy is the bluest in the 6C subsample.    

\subsection*{6C1019+39}

6C1019+39, at $z = 0.922$, is the brightest and lowest redshift galaxy
in the sample.  The combination of the images in the two HST filters
(Fig.~\ref{Fig: 6_5}) show that this galaxy is very red in colour,
especially when compared to the two other sources observed in more
than one HST filter (6C1217+36 and 6C1257+36).  The host galaxy
appears to be slightly elongated, and no obvious aligned structures
can be observed.  No radio core has been observed for this source. The
position of the host galaxy relative to the radio emission suggests
that the radio source is bent, with the radio jets lying on two
different axes, the southern jet being directed roughly due south and
the northern jet being directed roughly north east.  The southern lobe
contains an additional hotspot.  6C1019+39 is one of the smaller
galaxies in the $z \sim 1$ subsample, with a projected size of 67kpc.
The optical (rest--frame UV) spectrum of this source displays a high
velocity emission line component, and is well explained with radio
source shocks as the dominant ionization mechanism. The high velocity
component cannot be clearly linked to any optical structure, and is
offset to the north east of the peak of the continuum emission by less
than 1 arcsec.

\begin{figure*}
\vspace{4.2 in}
\begin{center}
\includegraphics{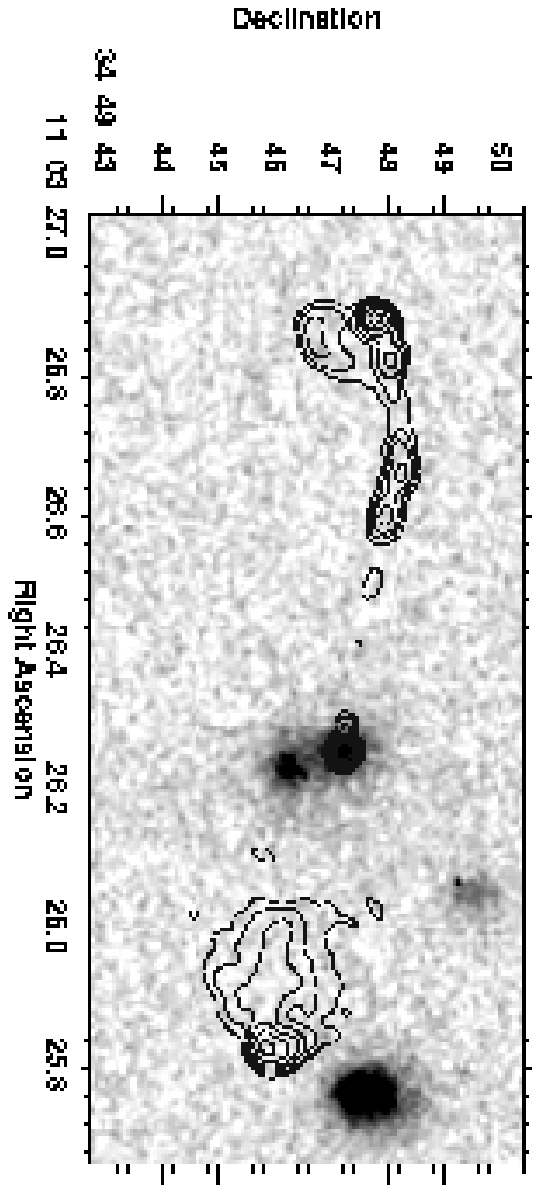}
\includegraphics{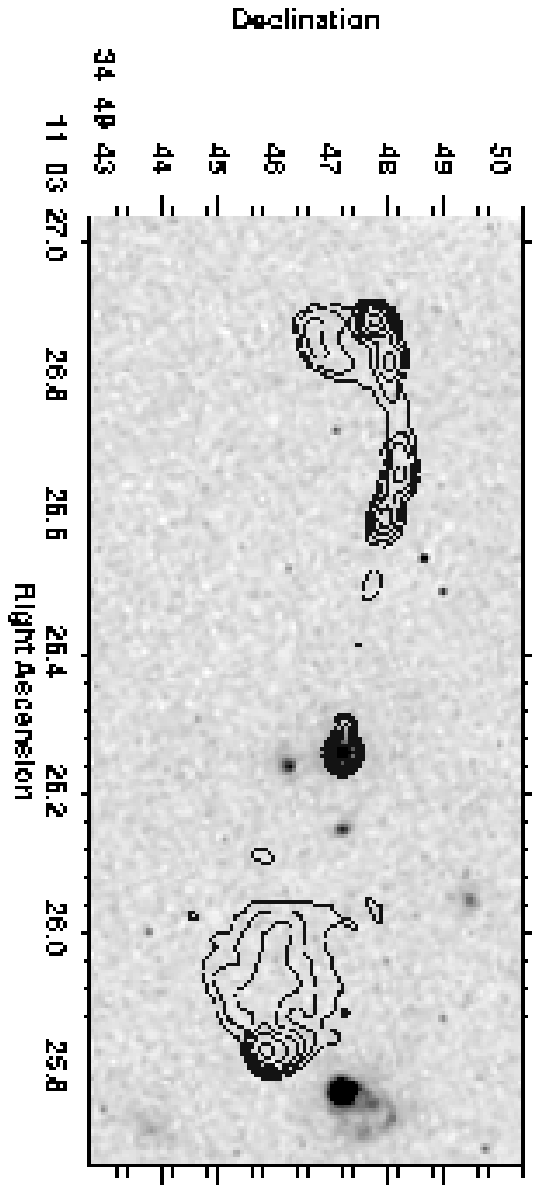}
\includegraphics{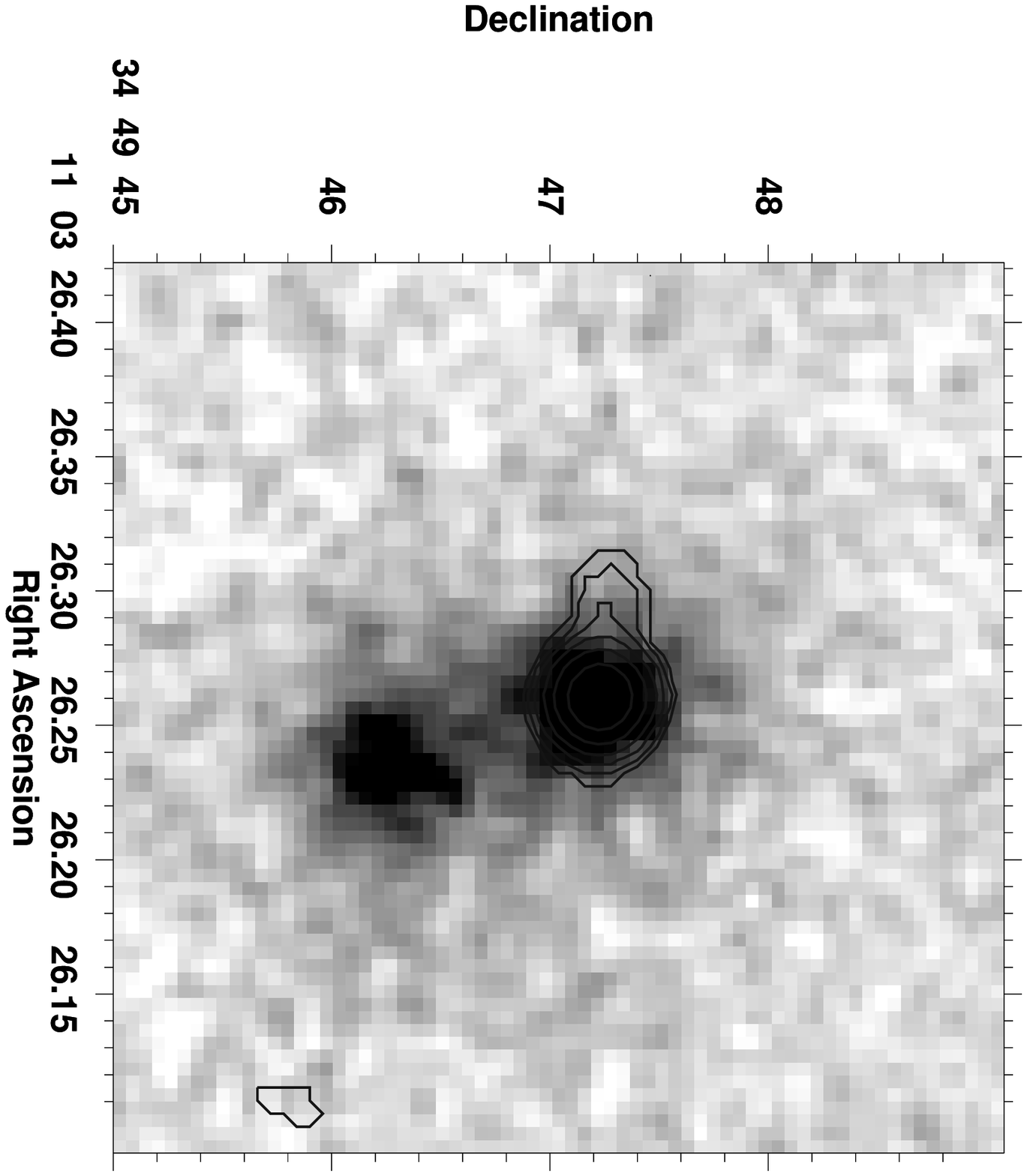}
\includegraphics{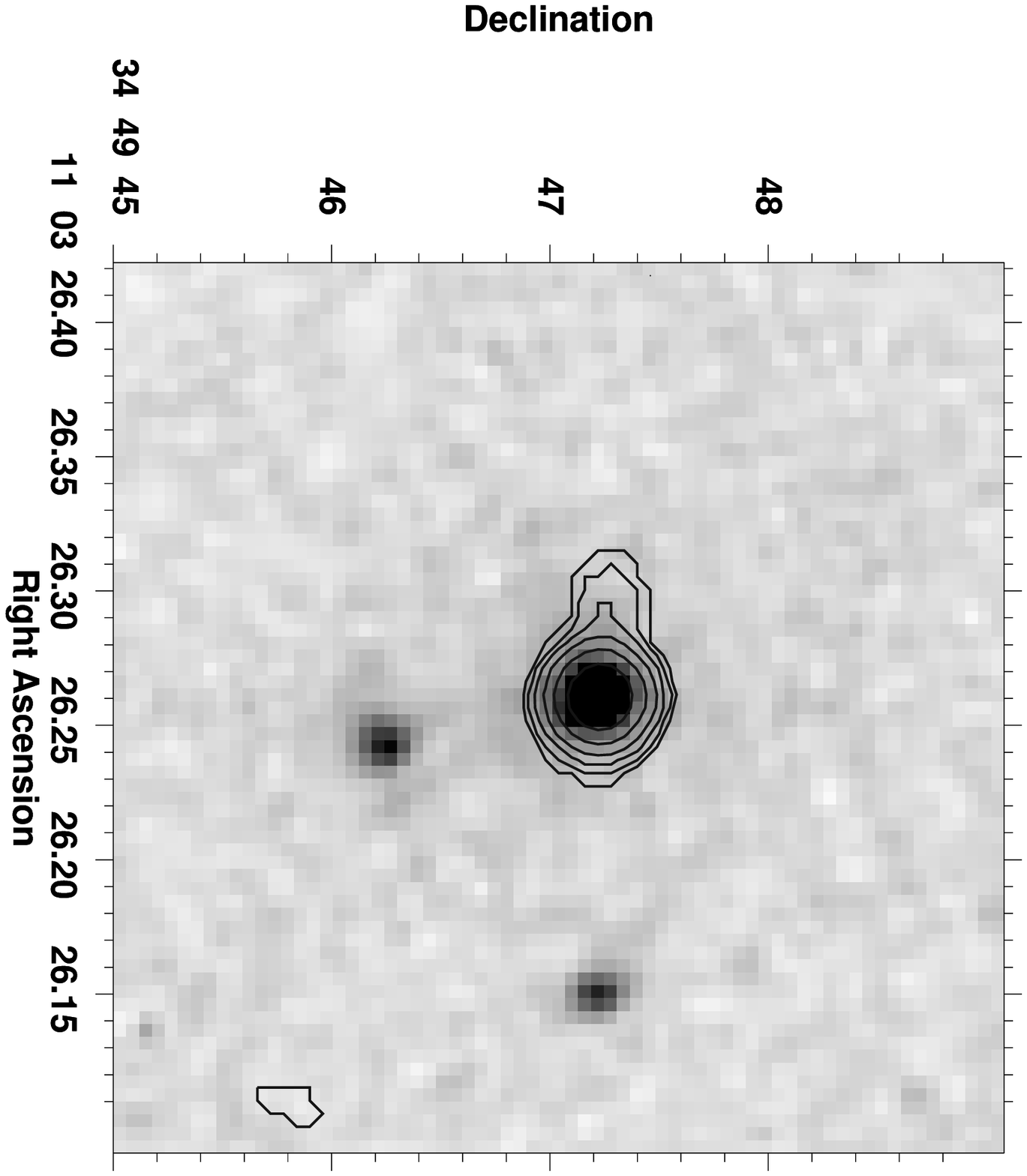}
\end{center}
\caption{6C1100+35: (a - top left) $K$-band image of 6C1100+35. (b - top
right) zoomed in $K$-band image of the central region of 6C1100+35. (c -
bottom left) HST F814W image of 6C1100+35. (d - bottom right) zoomed in HST
F814W image of the central region of 6C1100+35. Contour lines for all
images represent the 5GHz VLA  observation of this source, with contours at 100$\mu\rm{Jy}$ beam$^{-1} \times$ (-1,1,2,4...1024).  All
coordinates are in epoch J2000.0.  
\label{Fig: 6_6}}
\end{figure*}

\subsection*{6C1100+35}

6C1100+35, at $z = 1.44$, is one of the three higher redshift sources
in the sample. Fig.~\ref{Fig: 6_6} shows that this galaxy does not
show any noticeable aligned emission, although it 
does have a close companion galaxy $\sim 1^{\prime\prime}$ to the south.  
The F814W$-K$ colour for 6C1100+35 is redder than average for the
sample, but in keeping with the colours of other higher redshift 6C
and 3CR sources.   The fraction of the total radio flux observed in the bright
core is considerably higher than average for the sample.  A powerful
jet is also observed for this source, and the radio structure of 6C1100+35 is
very similar to that of a quasar.  However, this source has not been
classed as a quasar, due to the lack of any broad lines in its
spectrum (Rawlings, Eales \& Lacy 2001). The radial profile of the HST
emission (see Paper 2 for the full analysis of the radial profiles)
suggests a de Vaucouleurs profile, possibly with a very minor
contribution from an AGN at small radii. In addition the $K$-band
emission is clearly extended. The average colours of this
source support the interpretation that it is a radio galaxy rather
than a quasar.

\subsection*{6C1129+37}

\begin{figure*}
\vspace{5.4 in}
\begin{center}
\includegraphics{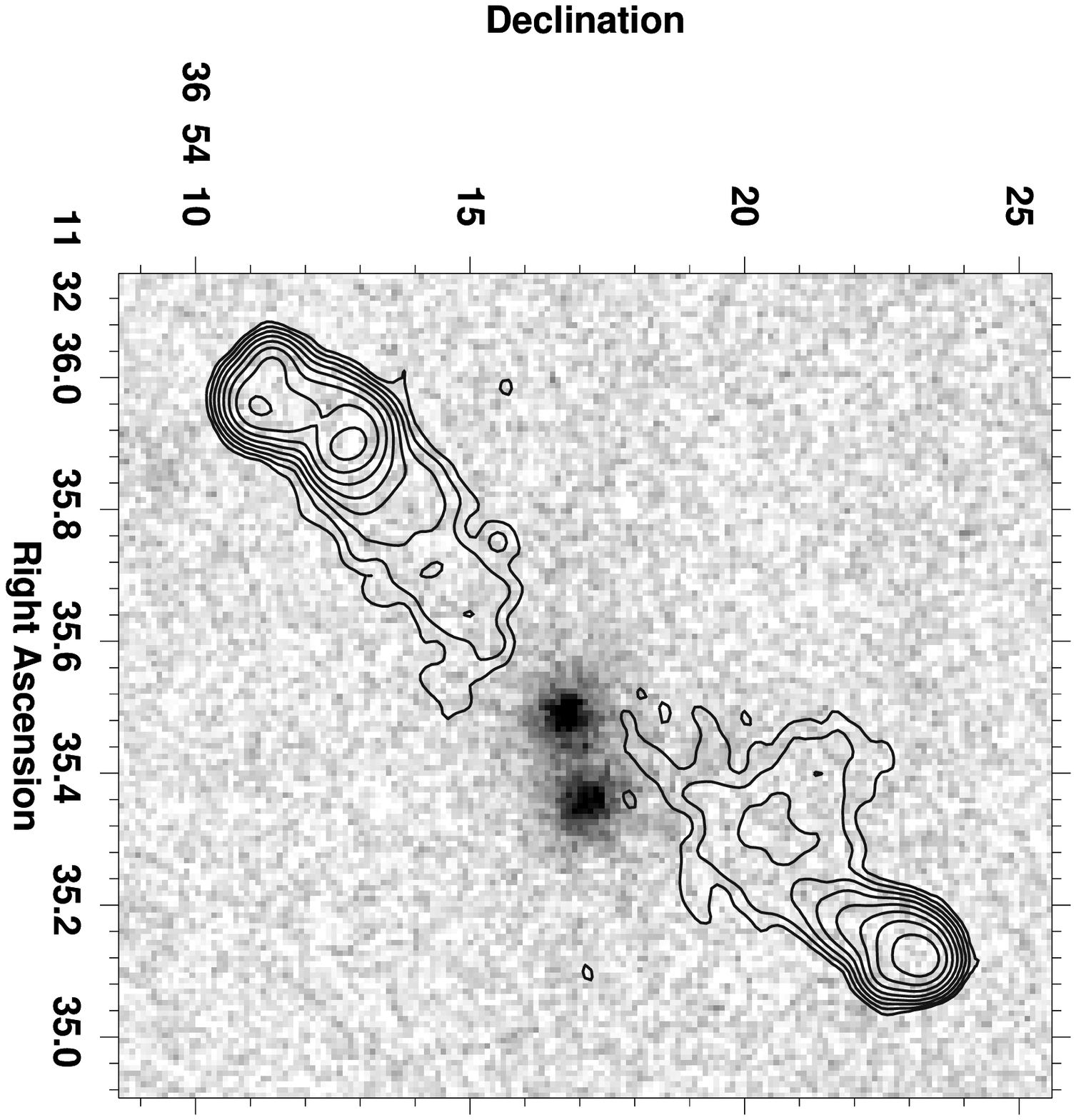}
\includegraphics{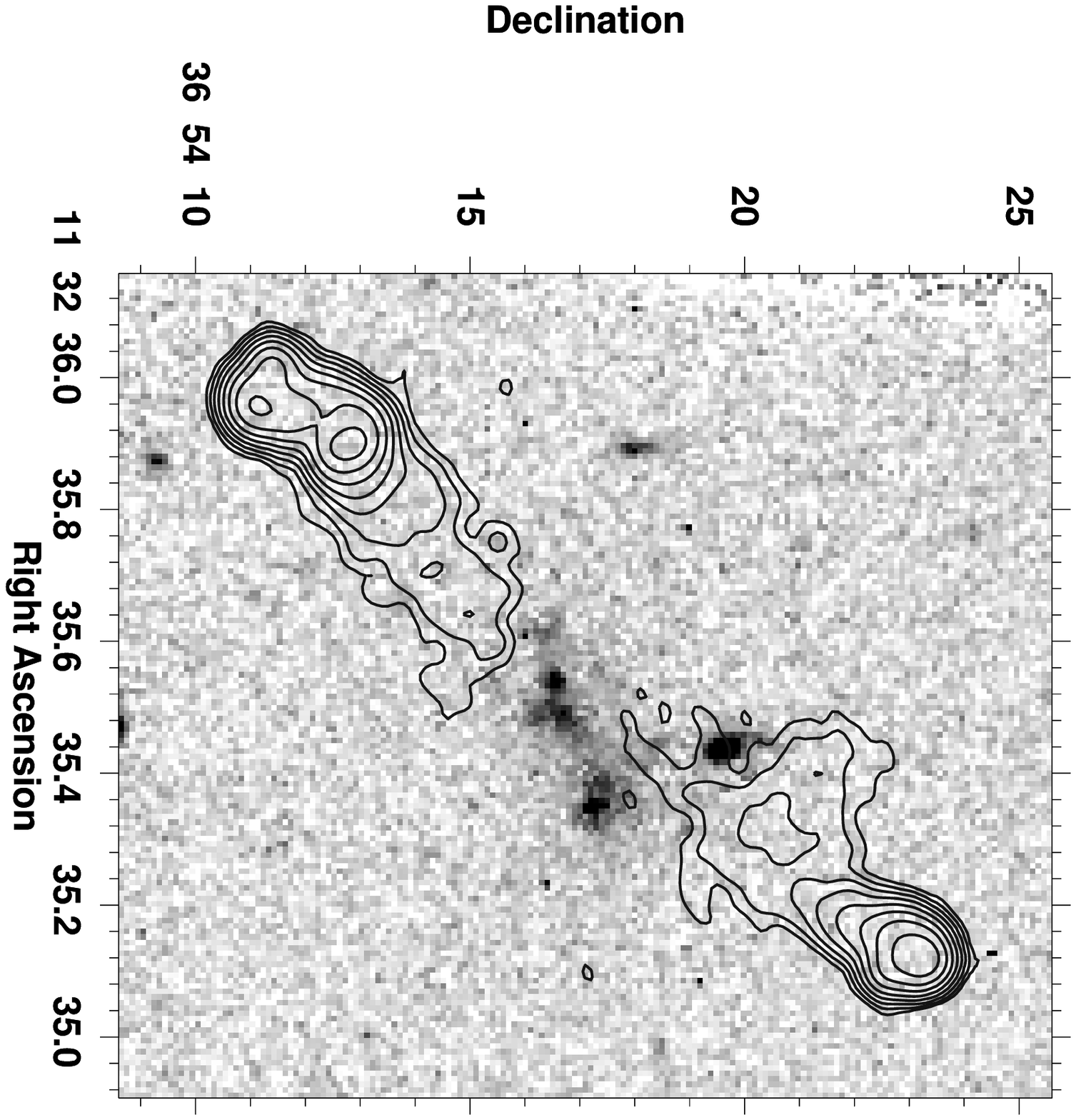}
\includegraphics{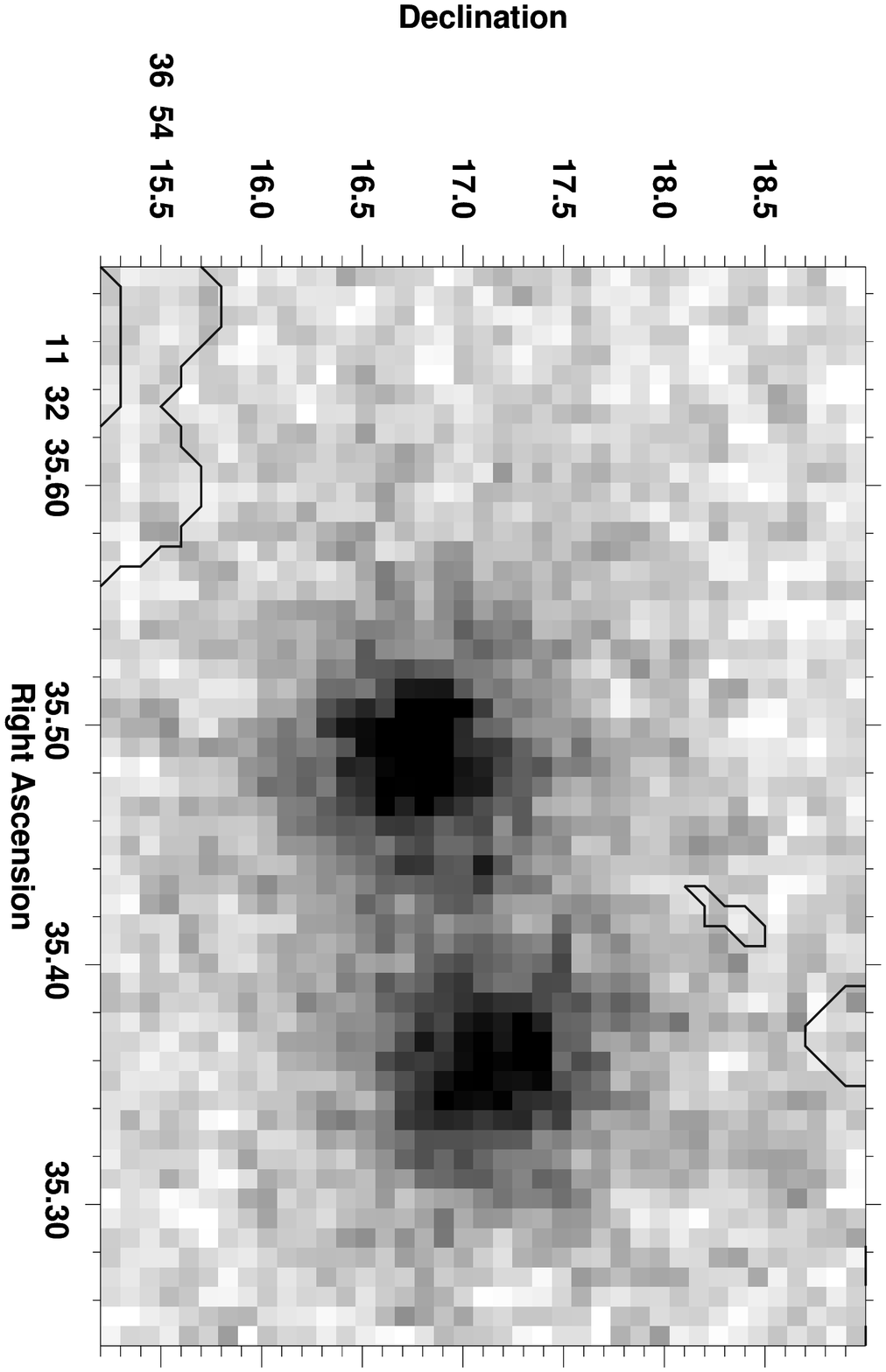}
\includegraphics{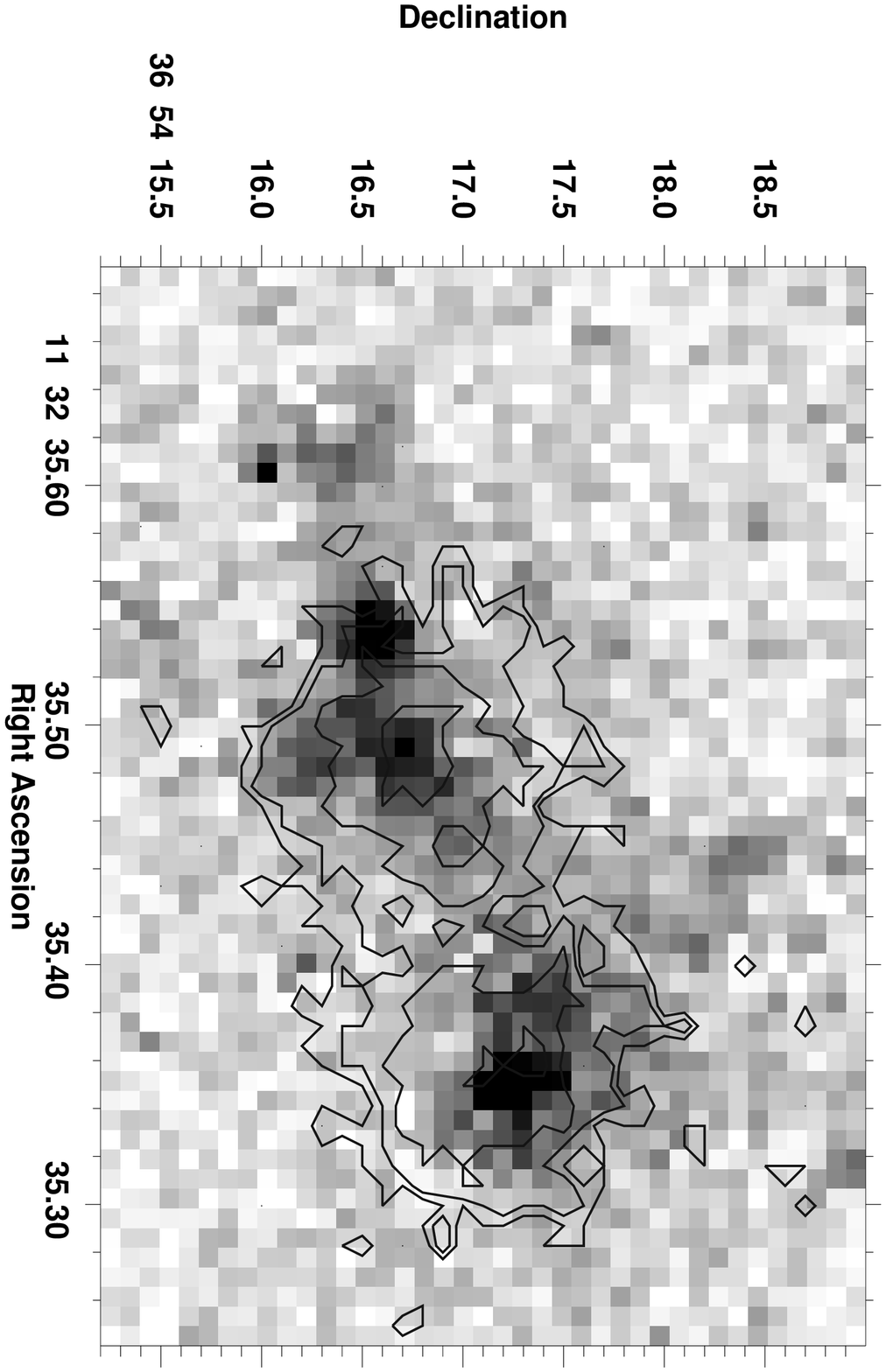}
\end{center}
\caption{6C1129+37: (a - top left) $K$-band image of 6C1129+37, with VLA
5GHz radio contours at 70$\mu\rm{Jy}$ beam$^{-1} \times$
(-1,1,2,4...1024). (b - top right) HST F702W image of 6C1129+37, 
with VLA 5GHz radio contours.  (c - bottom left) zoomed in $K$-band image
of the central region of 6C1129+37, with VLA 5GHz radio contours.  (d
- bottom right) zoomed in HST F702W image of the central region of
6C1129+37, with UKIRT $K$-band contours.   All coordinates are in epoch J2000.0.  
\label{Fig: 6_7}}
\end{figure*}

6C1129+37, at a redshift of $z = 1.060$, is a very interesting radio
galaxy.  Infrared imaging (Fig.~\ref{Fig: 6_7}) shows two elliptical
galaxies of a similar 
size in close proximity.  The radio source jet passes very close to
the companion galaxy (W) of the radio source (E).  The HST image shows
a number of bright knots and extended features surrounding both
galaxies, none of which are visible in the infrared images. It is
quite likely that these galaxies are interacting with each other.
On the HST image, the radio source is identified with the second
brightest eastern knot, and the companion galaxy with the brightest
western knot.  This is illustrated by Fig.~\ref{Fig: 6_7}d, where the
UKIRT contours clearly identify the location of the galaxies on the
HST image.  Although no radio core has been observed for this source,
the galaxy identified with the radio source is coincident with the
peak of the line emission, and lies more directly between the two
radio lobes.  The south eastern lobe contains three hotspots, and is
more strongly polarized than the northern radio lobe. 
 
The extensive emission surrounding both galaxies is not dominated by
line emission, unlike 6C0943+39.  The line emission from this source
is most important close to the radio galaxy. The spectra of this
source suggest that shock ionization is not an important factor, but
AGN photoionization does not predict the line ratios well either,
suggesting that a complex combination of ionization mechanisms may be
at work.  The 
gas surrounding the radio galaxy displays a very steep velocity
gradient, whereas that surrounding the companion galaxy is at a
roughly constant relative velocity (Inskip et al 2002a).
 
The measured F702W$-K$ for this galaxy is very blue, even excluding
the bright blue feature to the north which may perhaps be
associated with the radio galaxy.   This blue
colour is due to the extensive aligned emission surrounding both
galaxies observed on the HST image.  The $J-K$ colour is fairly
average for the subsample, however.  The companion galaxy is
fractionally fainter in $K$, with a bluer value for F702W$-K$.  In
places on the HST image, the extended emission is brighter than the
radio galaxy it surrounds.  The morphology of the excess UV emission
surrounding this source is the most extreme observed in the 6C
subsample.
\begin{figure*}
\vspace{4.9 in}
\begin{center}
\includegraphics{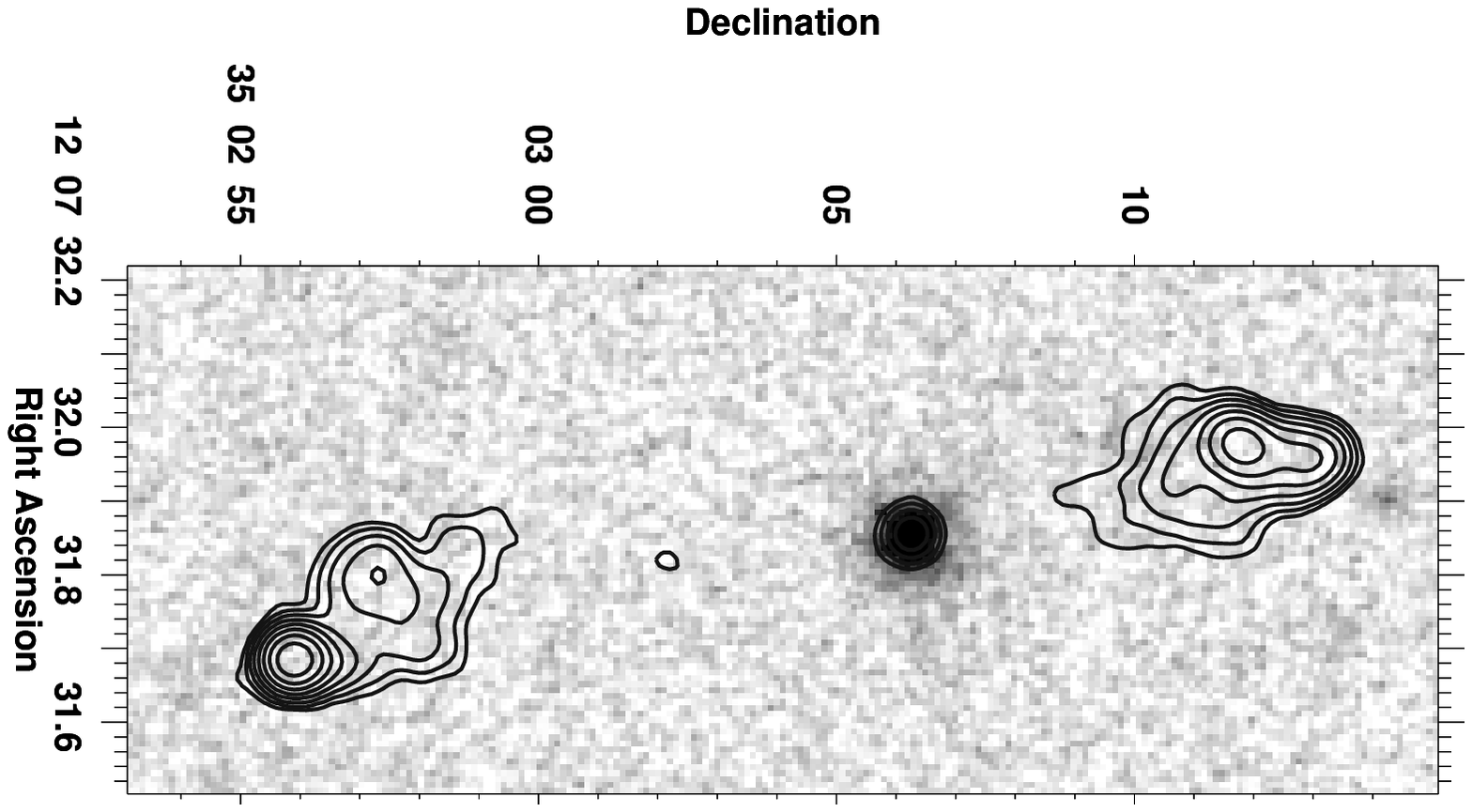}
\includegraphics{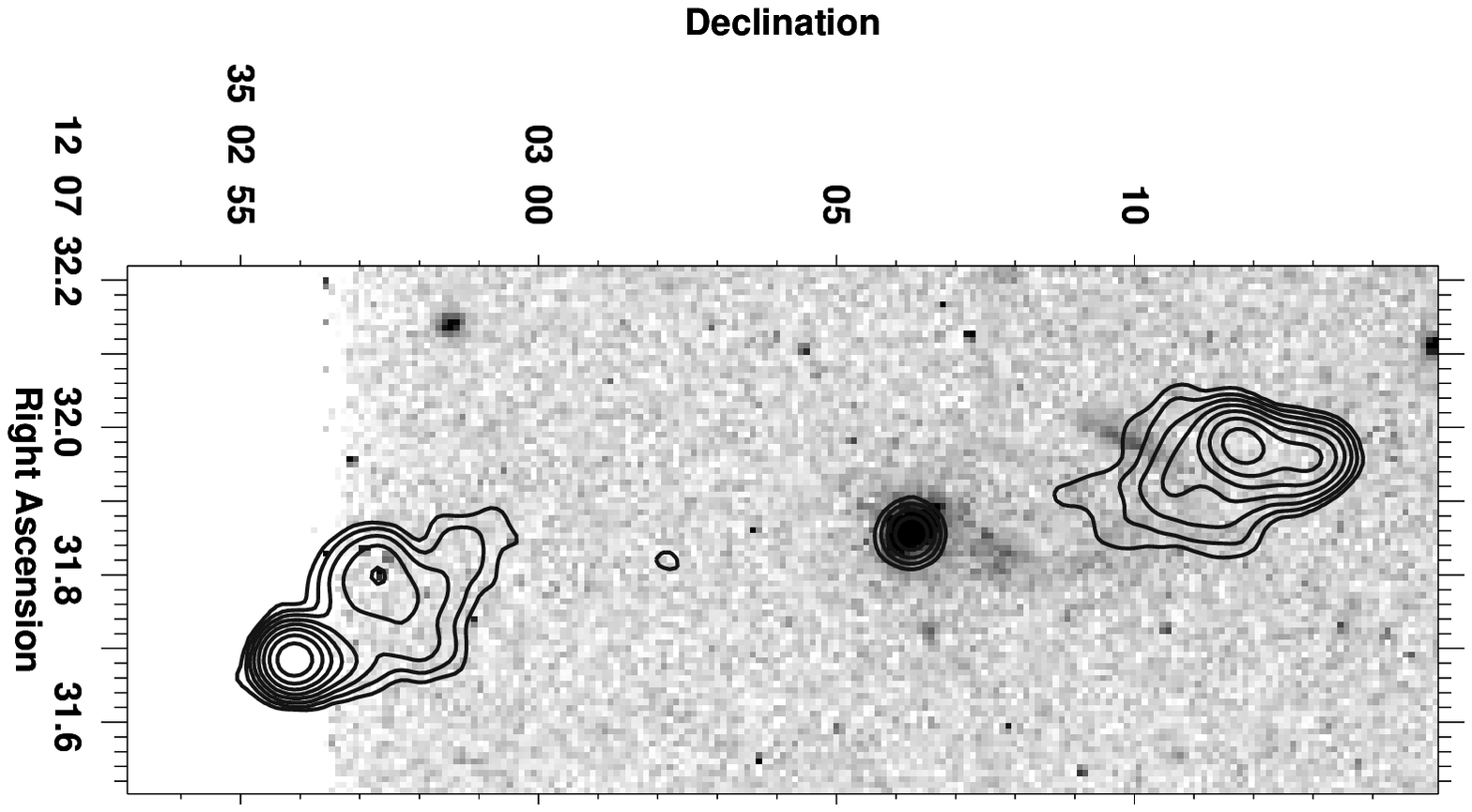}
\end{center}
\caption{6C1204+35: (a - left) $K$-band image of 6C1204+35. (b - right)
HST F814W image of 6C1204+35.  Contour lines for both images represent the
5GHz VLA observation of this source, with contours at 150$\mu\rm{Jy}$ beam$^{-1} \times$ (-1,1,2,4...1024).  All coordinates are in epoch J2000.0.
\label{Fig: 6_8}}
\end{figure*}

\begin{figure}
\vspace{8.1 in}
\begin{center}
\includegraphics{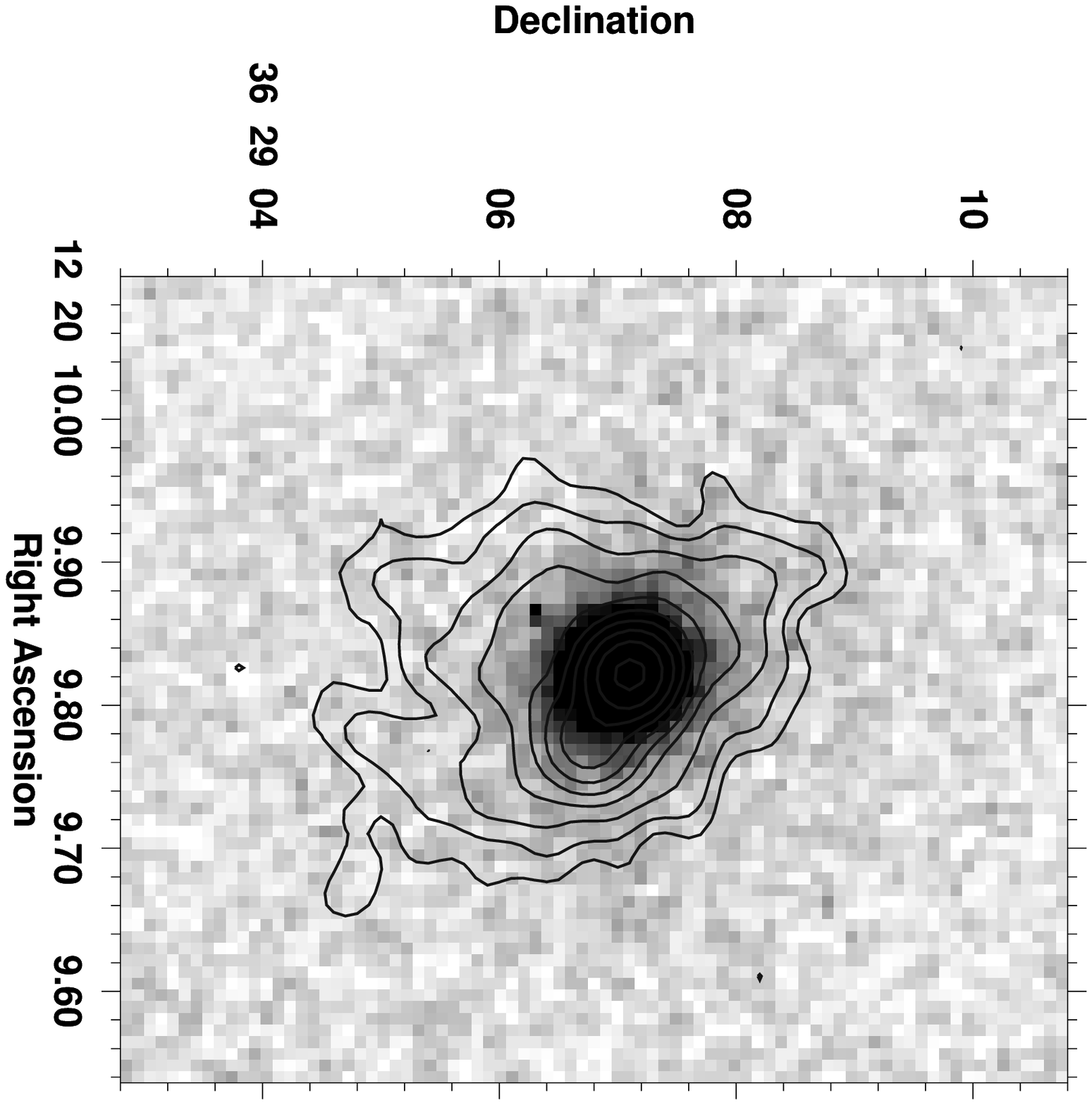}
\includegraphics{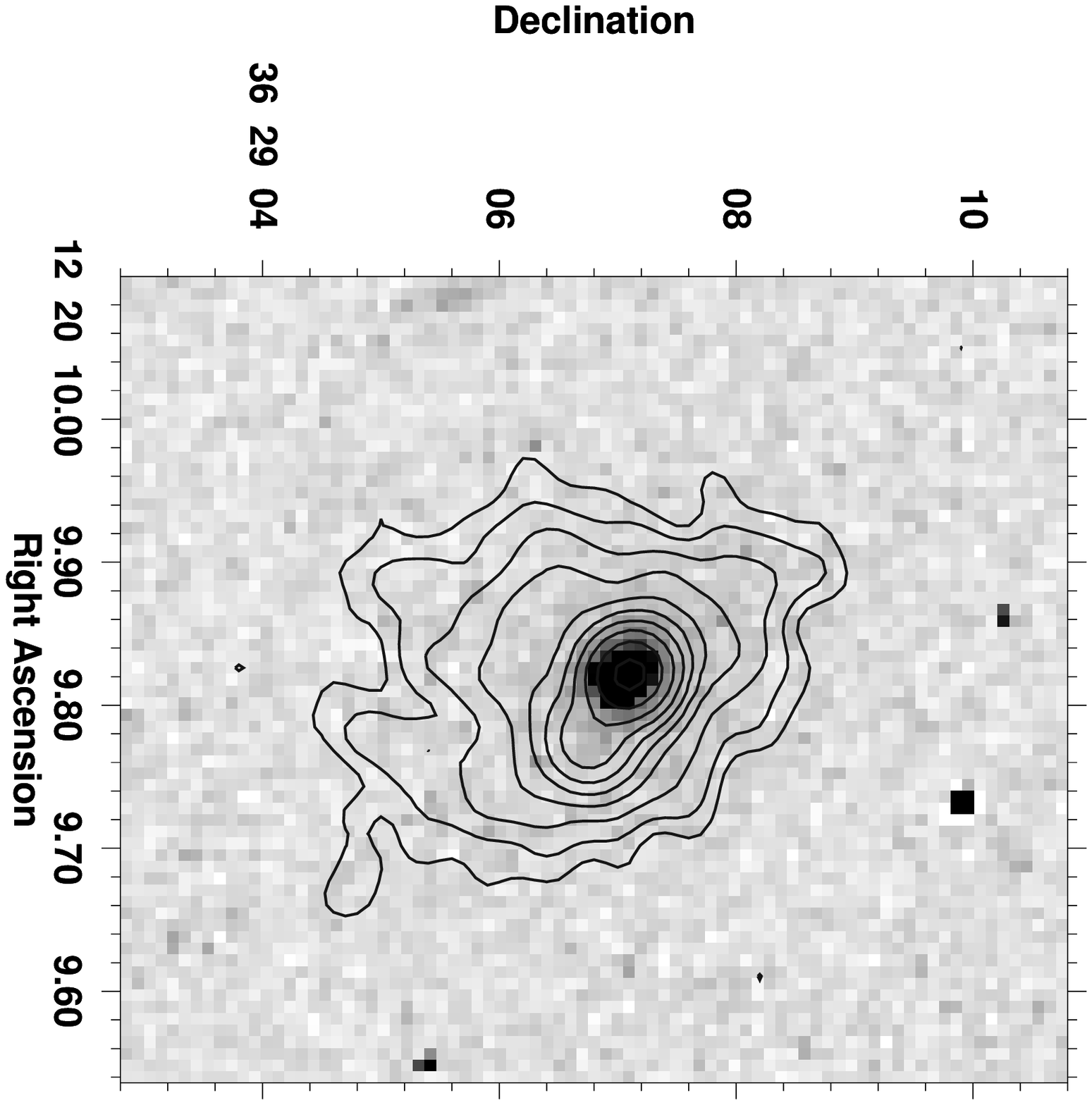}
\includegraphics{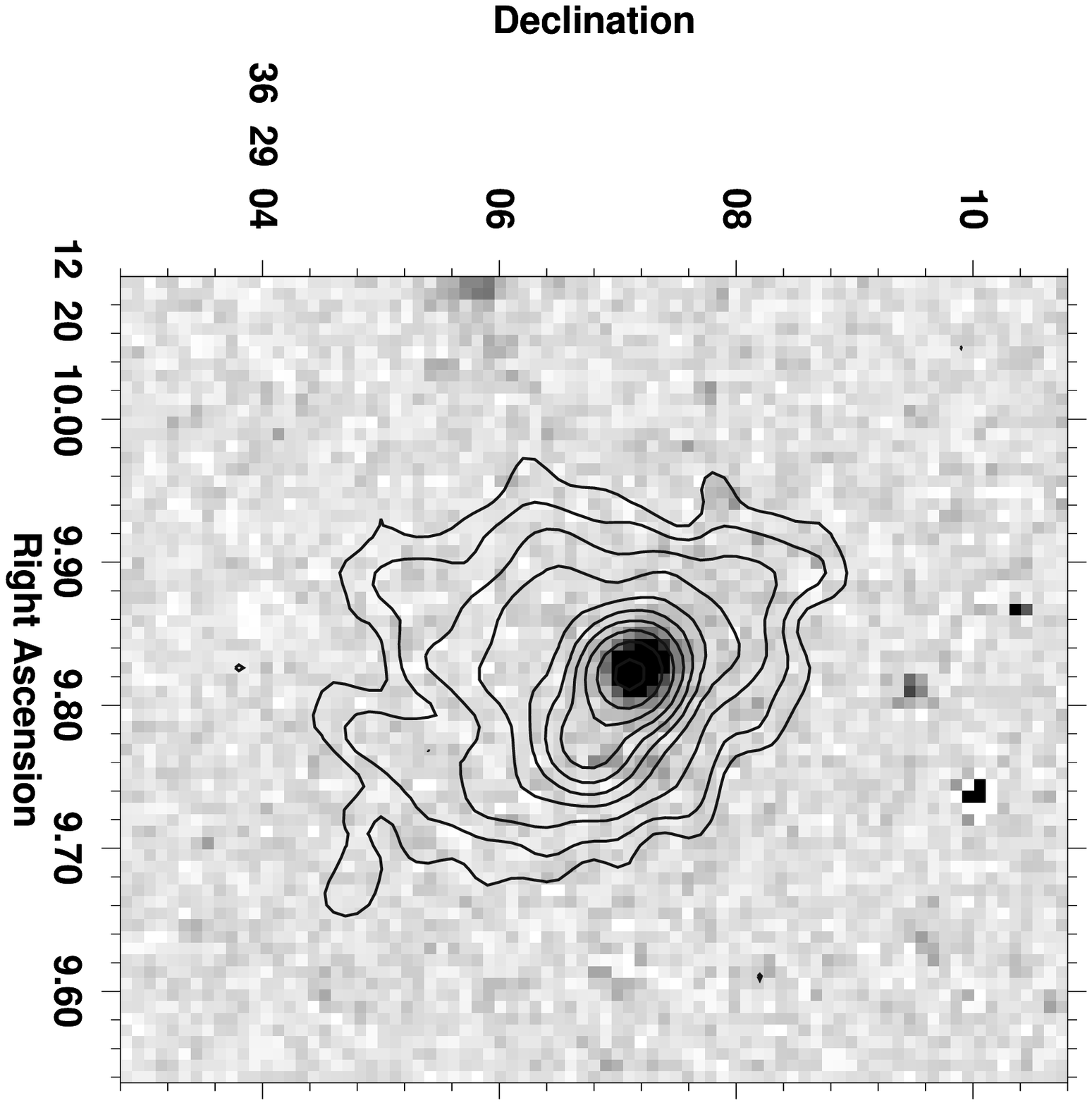}
\end{center}
\caption{6C1217+36: (a - top) $K$-band image of 6C1217+36. (b - 
centre) HST F814W image of 6C1217+36.  (d - bottom) HST F606W image of
6C1217+36.  Contour lines for all images represent the 5GHz VLA 
observation of this source, with contours at 150$\mu\rm{Jy}$ beam$^{-1} \times$ (-1,1,2,4...1024).  All coordinates are in epoch J2000.0. 
\label{Fig: 6_9}}
\end{figure}

\subsection*{6C1204+35}

6C1204+35, at $z = 1.37$, is the third high redshift source in the
sample.  The colours calculated for this source are average for its
redshift.  There is some doubt as to the origin of the aligned
emission visible on the HST image (Fig.~\ref{Fig: 6_8}), which could perhaps be a tidal arc
associated with a faint spiral companion to the north (12 07 32.0 +35
03 10).  Alternatively,
both the arc and the possible spiral companion could simply be extended emission features related to the
AGN activity.  Line emission and nebular continuum emission are
clearly important contributors to the total emission from this source
(cf. Table 3), although new deep spectroscopic
observations are not available for this source.  Therefore, it is
unknown whether AGN photoionization
or ionization by shocks is the dominant ionization
mechanism.

\subsection*{6C1217+36}

6C1217+36 ($z = 1.088$) does not appear to be a standard FRII radio
source, and is an anomalous galaxy in the sample.  Radio observations
show a compact double source in the centre, surrounded by a more
extensive halo radio emission. Unusually, the central regions of this
source are not depolarized, as would be expected for radio emission
from well within the host galaxy.  The imaging observations
(Fig.~\ref{Fig: 6_9}) simply show a bright elliptical galaxy, with no
trace of any aligned emission. However, the radial profile of the
HST emission (analysed in full in Paper 2) seems to show evidence for
a point source contribution, particularly in the F606W filter.
Although this galaxy has a similar F606W-F814W colour to 6C1257+36
(which has also been observed in two 
HST filters, and is at a similar redshift), the arc of aligned
emission present in the images of 6C1257+36 is brightest in the F606W
filter, suggesting that of these two sources 6C1217+36 has a bluer
host galaxy in this colour.  However, the F814W$-K$ colour for this
source is fairly red, consistent with that of a passive elliptical
galaxy at this redshift.  Deep spectroscopic observations show that
this source has very weak line emission. 

\subsection*{6C1256+36}

\begin{figure*}
\vspace{6.2 in}
\begin{center}
\includegraphics{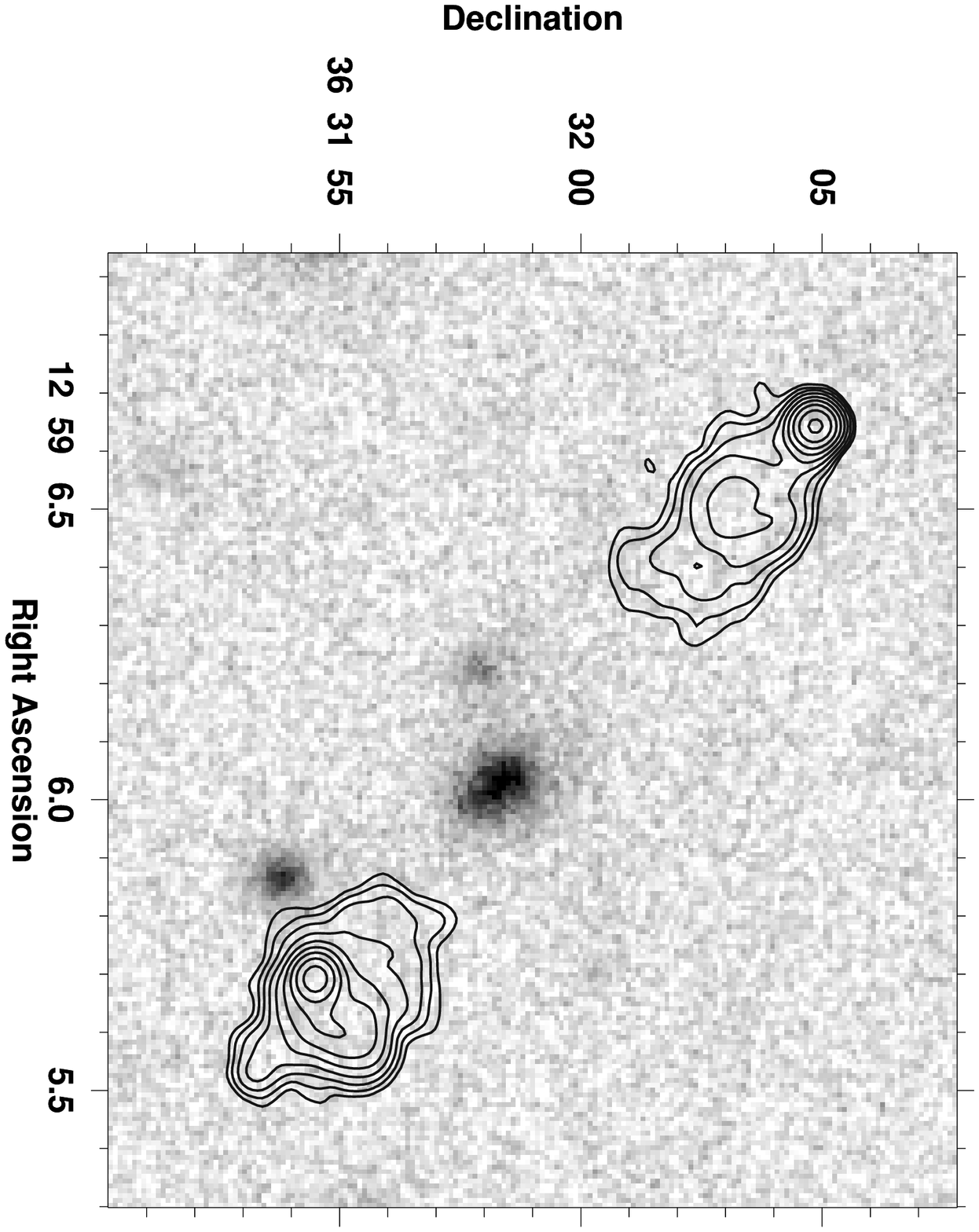}
\includegraphics{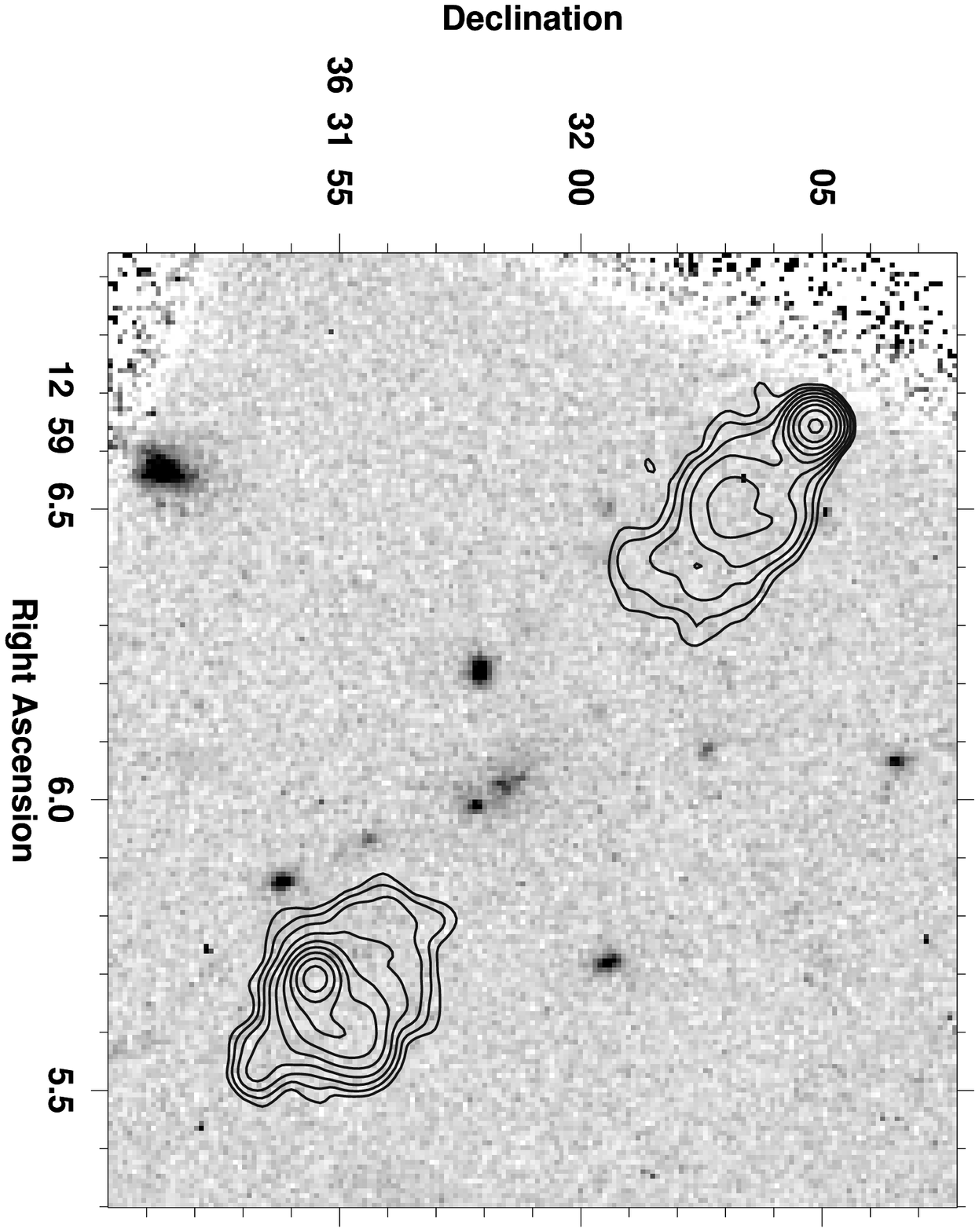}
\includegraphics{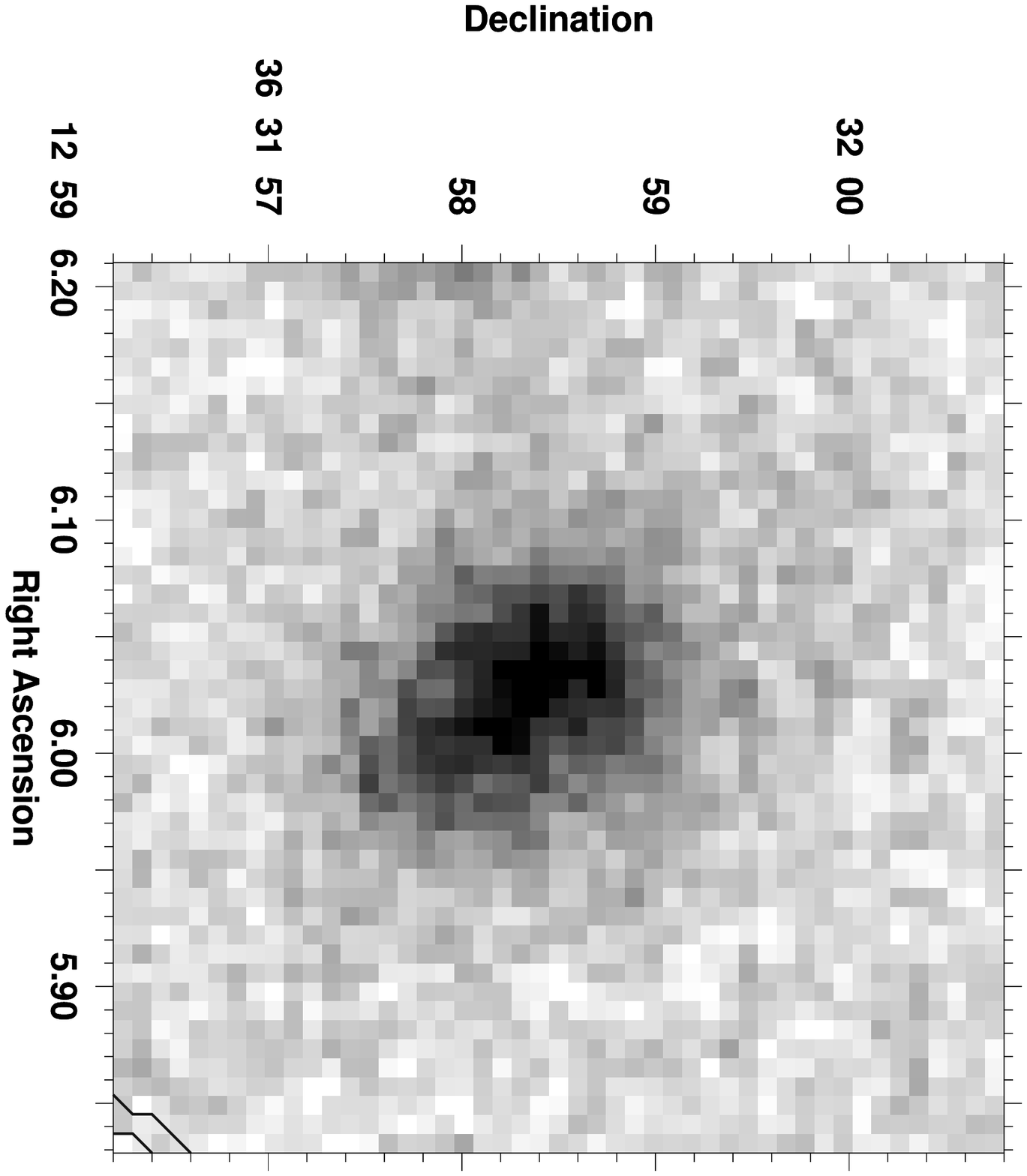}
\includegraphics{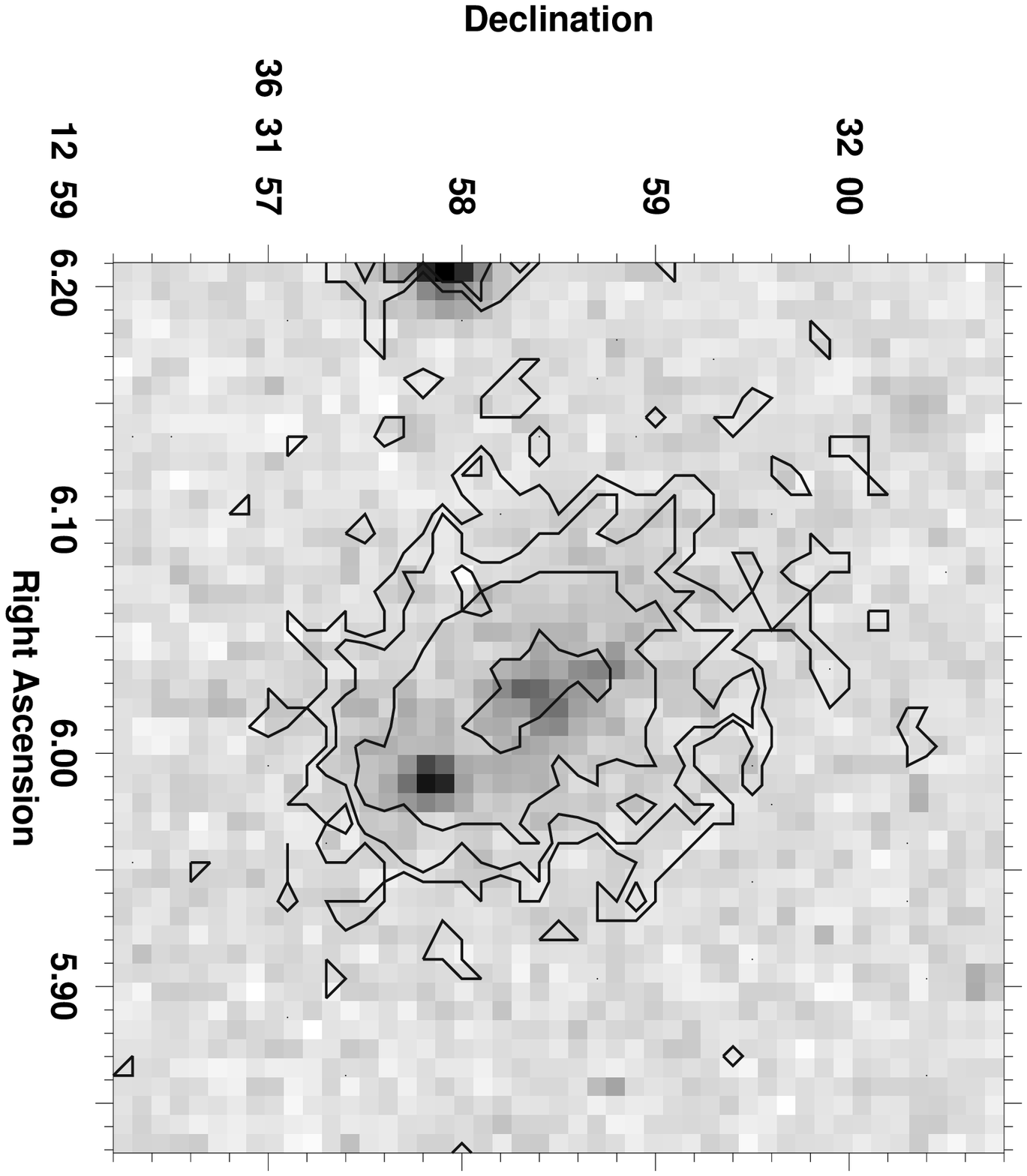}
\end{center}
\caption{6C1256+36: (a - top left) $K$-band image of 6C1256+36, with VLA
5GHz radio contours at 100$\mu\rm{Jy}$ beam$^{-1} \times$ (-1,1,2,4...1024). (b - bottom left) HST F702W image of 6C1256+36,
with VLA 5GHz radio contours. (c - top right) zoomed in $K$-band image
of the central region of 6C1256+36.  (d - bottom right) zoomed in HST
F702W image of the central region of 6C1256+36, with UKIRT $K$-band
contours.  All coordinates are in epoch J2000.0.   
\label{Fig: 6_10}}
\end{figure*}

The UKIRT images of 6C1256+36 ($z = 1.128$) show a seemingly elongated
elliptical galaxy (Fig.~\ref{Fig: 6_10}). With UKIRT $K$-band contours
overlaid on the HST 
F702W image, it is seen that the brightest component of the HST emission
does not lie at the peak of the $K$-band emission. The $K$-band image is
also extended in this region. The combined HST and UKIRT
observations of this source show that this extended emission region is bluer
than the elliptical host galaxy of the radio source, and is most
likely due to the alignment effect. Roche, Eales \&
Hippelein  (1998) found that 6C1256+36 exists in a richer environment
than average for radio galaxies at $z \sim 1$, and claim that it lies
within an Abell class 2 cluster.  The $J-K$ and F702W$-K$ colours for
this source are average for the sample.  Spectroscopic observations of
this source suggest that both shock ionization and AGN photoionization
are important.  No radio core has been observed for this source.  

\subsection*{6C1257+36}

\begin{figure*}
\vspace{6.7 in}
\begin{center}
\includegraphics{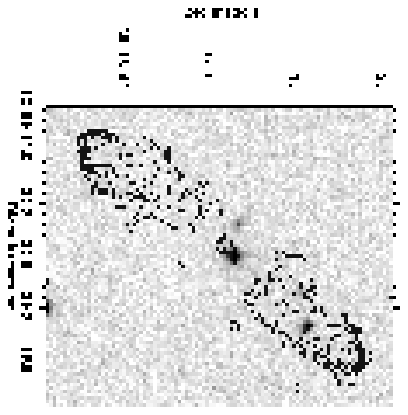}
\includegraphics{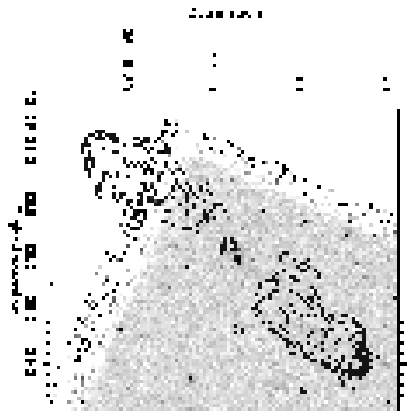}
\includegraphics{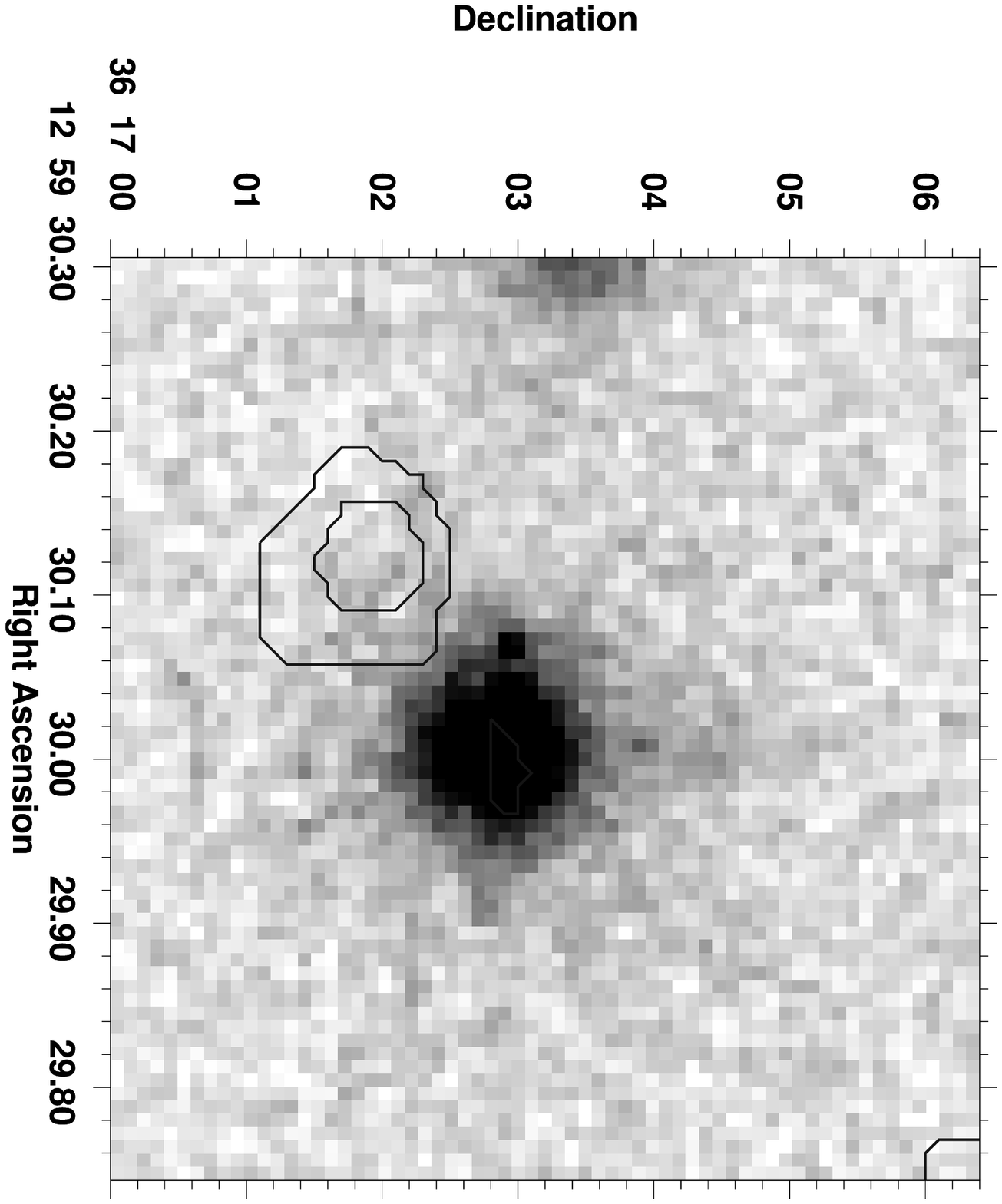}
\includegraphics{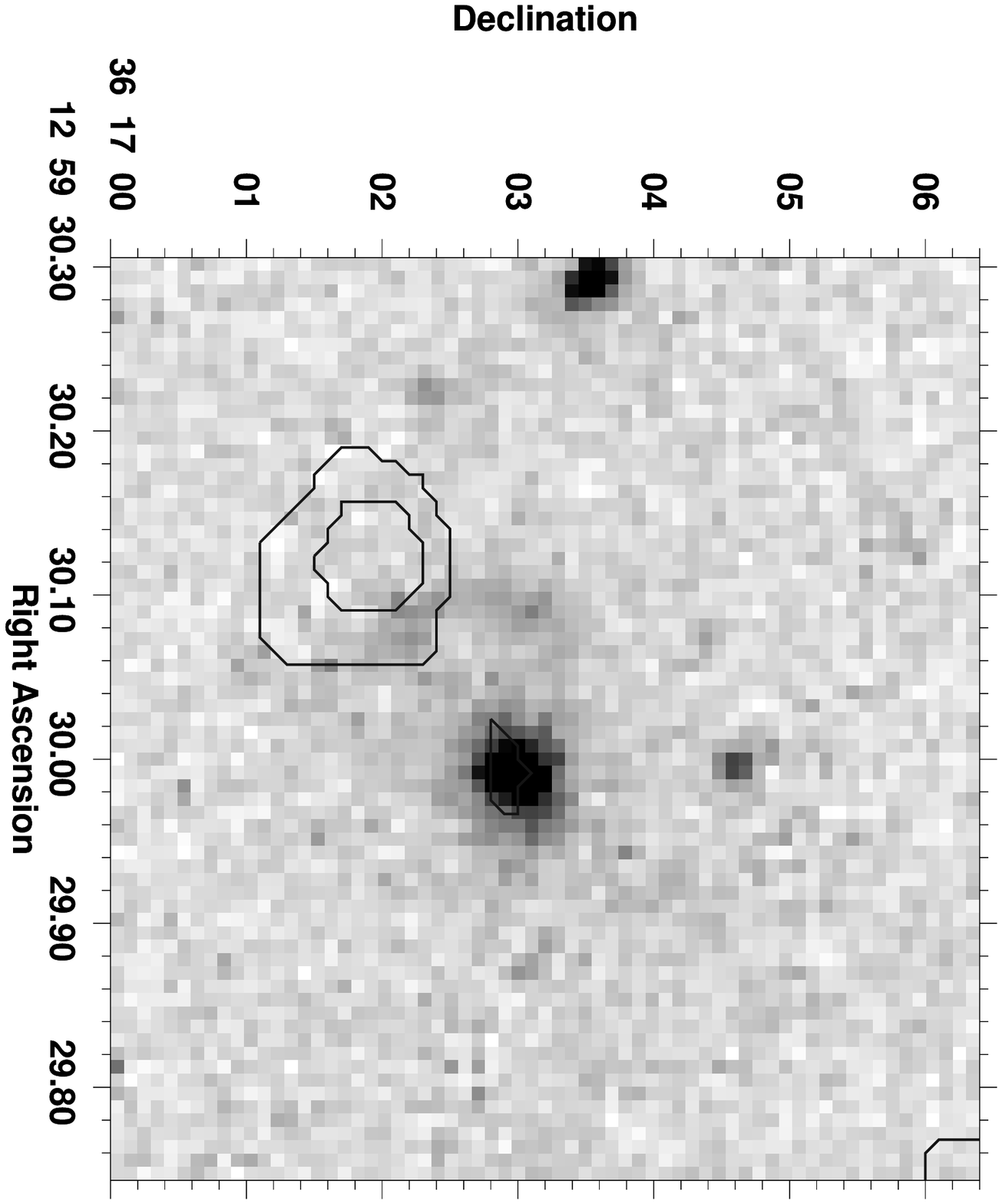}
\includegraphics{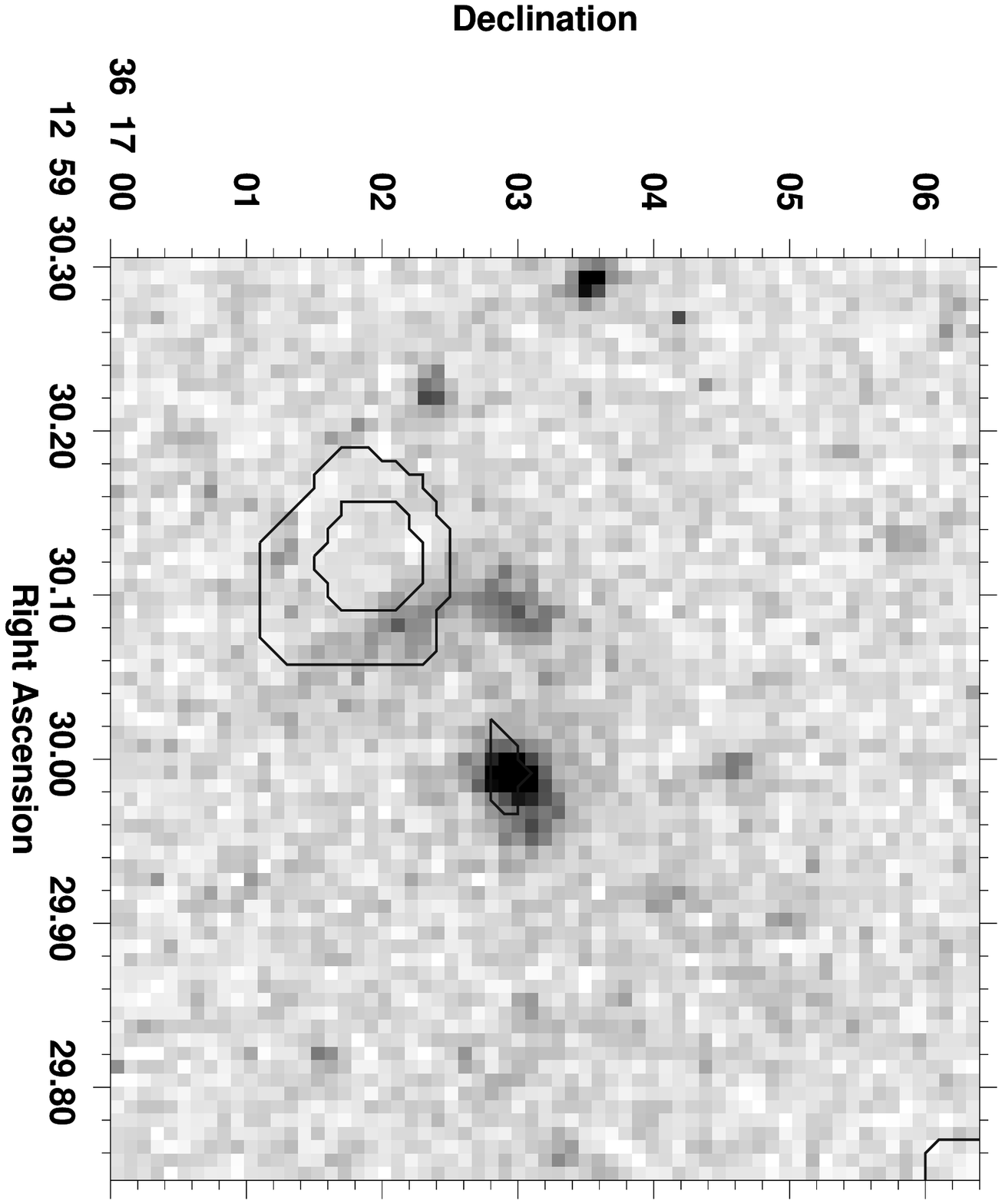}
\end{center}
\caption{6C1257+36: (a - top left) $K$-band image of 6C1257+36. (b - 
bottom left) HST F606W image of 6C1257+36. (c - top right) zoomed in
$K$-band image of 6C1257+36. (d - centre right) zoomed in HST F814W
image of 6C1257+36.  (e - bottom right) zoomed in HST F606W image of
6C1257+36.  Contour lines for all images represent the 5GHz VLA
observation of this source, with contours at 80$\mu\rm{Jy}$ beam$^{-1}
\times$ (-1,1,2,4...1024).  All coordinates are in epoch J2000.0.  
\label{Fig: 6_11}}
\end{figure*}

The radio contours for 6C1257+36 ($z = 1.004$) show two possible core
candidates, of which the NW knot has been found to be coincident with
the host galaxy (Fig.~\ref{Fig: 6_11}).  Best {\it et al} (1999) correctly identified the
fainter, NW knot as the radio core due to its flatter spectral index.  
In addition, the south eastern lobe displays a double hotspot and
little polarized emission.
The HST imaging of 6C1257+36 in two filters shows that this radio galaxy
has a fairly red elliptical host galaxy, consistent with an old
stellar population, with a blue arc of
aligned emission.    The blue arc appears to be nearly coincident with
the SE knot in the radio emission.  A further small region of
bluer emission is observed to the northwest, again aligned with the
radio axis. 
 
The spectra of this source show that the line emission is well
explained by photoionization by an obscured AGN. However, the large
physical extent of the line emission and its extreme kinematic
properties suggest that the line emitting regions in this source have
been influenced by the radio knot, which may have recently passed
through the gas clouds forming the arc.
The F814W $-$ F606W colour for this galaxy is comparable to that of
6C1217+36, and considerably bluer than that of 6C1019+39. However,
this source displays extensive blue aligned emission whereas 6C1217+36
does not.  Nebular continuum emission is significant in both filters
whereas line emission is only significant in the redder of the two
filters; the blue arc is clearly an important continuum emission
feature associated with an alternative emission mechanism.

\section{Discussion}

The initial results of the imaging observations and a preliminary
analysis of the 
galaxy colours have provided a number of interesting results. 
By combining the results of the imaging observations with previous
deep spectroscopic observations (Inskip et al 2002a) we have been able
to determine the contribution of several different emission mechanisms
to the aligned emission. Line emission and nebular continuum emission are
present in the wavelength range sampled by the HST observations, in
similar proportions to that found in observations of 3CR sources at
the same redshift.  Whilst these processes are important in the
diffuse extended emission surrounding 6C0943+39 (contributing up to
50\% of the observed flux at a distance of $>$4\arcsec\ from the host
galaxy), the bright, blue knotty features observed around 6C1129+37
cannot be explained by regions of bright line emission. 
Overall, line emission and nebular continuum emission are
not the dominant processes producing the alignment effect in these
galaxies. 

In general, the magnitudes determined within the 4\arcsec\ and
9\arcsec\ diameter apertures are comparable.  Any differences can be
fully accounted for by the presence of very extended UV/optical
emission (e.g. 6C1129+37, 6C1204+35) and the extended nature of the
host galaxies, which typically have characteristic radii of
$\sim$1-3\arcsec\ (Best et al 1998; subsequent papers in this series
(in prep)).
For several sources, the galaxy colours are complicated by
the presence of point source contributions from the AGN. An accurate
quantification of the point source contributions and detailed
analysis of the galaxy colours, including a comparison with 3CR
sources in order to investigate the influence of radio power, has therefore
been deferred to a later paper in this series.  
Although a full morphological analysis of the galaxies is also
deferred to a later paper, several
conclusions can be reached simply on the basis of the observed
appearance of the galaxies.  
\begin{enumerate}
\item[$\bullet$] The infrared observations of the $z \sim 1$ 6C radio
galaxies do not display any significant extended emission aligned
along the radio axis.   A similar result was obtained for the $K$ band
observations of 3CR sources at the same redshift, confirming the
wavelength dependence of 
the alignment effect.   The slightly more elongated appearance of the
galaxy 6C1256+36 is most likely to indicate an elongated elliptical
morphology, rather than extended aligned
emission in the $K$-band or alternatively the presence of a close
companion galaxy.    
\item[$\bullet$] A number of sources appear to be interacting or
undergoing mergers.  These include 6C1011+36, previously
proposed to be associated with a rich Abell class 2 cluster (Roche,
Eales \& Hippelein 1998).  6C1129+37 is
clearly shown to be two merging ellipticals, rather than a single
galaxy with bright aligned infrared emission, as had previously been
considered. 6C1100+35 appears to have a close companion to the
south. 
\item[$\bullet$] A wide range of extended structures are observed in
the HST images of the 6C galaxies.  Some sources show considerable
extended emission, which is generally well aligned with the radio
source axis.   The variety of features observed are generally very
similar to those seen around 3CR sources at the same redshift, and
include bright knots, diffuse emission, linear features and arcs.  In
particular, arc-like aligned structures are observed around 6C1017+37
and 6C1204+35, reminiscent of those seen around 3C280 and 3C267.
Optically passive sources are also found in both samples
(e.g. 6C1019+39, 6C1217+36 and 3C65).  However there are some
features, such as the bright blue object to the 
north of 6C1129+37, which (if indeed associated with AGN activity) do not
resemble anything previously observed in the more powerful 3CR subsample.
\item[$\bullet$] As in the case of the 3CR $z \sim 1$ subsample,
extreme features were typically not observed around the larger radio
galaxies in the sample.  However, the strong trend observed in the 3CR
subsample for the smaller sources to display several bright knots of
aligned emission is not seen in the 6C data.  On the whole, the
aligned features observed around the 6C sources appear less luminous
and less extensive than those surrounding $z \sim 1$  3CR radio
sources.  Fewer bright emission components are observed, and these are
generally situated in closer proximity to the host galaxy.
\end{enumerate}

\section*{Acknowledgements}

KJI acknowledges the support of a PPARC research studentship and a
Lloyds Tercentenary Foundation Research Fellowship.  PNB is grateful
for the generous support offered by a Royal Society Research 
Fellowship. The United Kingdom Infrared Telescope is operated by the
Joint Astronomy Centre on behalf of the U.K. Particle Physics and
Astronomy Research Council. Some of the data reported here were
obtained as part of the UKIRT Service Programme.
Parts of this research are based on observations made with the
NASA/ESA Hubble Space Telescope, obtained at the Space Telescope
Science Institute, which is operated by the Association of
Universities for Research in Astronomy, Inc., under NASA contract NAS
5-26555. These observations are associated with proposals \#6684 and
\#8173.

\label{lastpage}

\end{document}